\pdfoutput=1
\RequirePackage{ifpdf}
\ifpdf 
\documentclass[pdftex]{sigma}
\else
\documentclass{sigma}
\fi

\numberwithin{equation}{section}

\newtheorem{Theorem}{Theorem}[section]
\newtheorem*{Theorem*}{Theorem}

\theoremstyle{definition}

\newtheorem{Remark}[Theorem]{Remark}

\usepackage{tikz,tikz-cd}
\usepackage{tikz-feynman}
\usetikzlibrary{decorations.pathreplacing}
\usetikzlibrary{calc,tikzmark}

\def\cE{\mathcal E}\def\cG{\mathcal G}
\def\cL{\mathcal L}
\def\cN{\mathcal N}\def\cO{\mathcal O}

\def\CC{\mathbb C}

\def\RR{\mathbb R}

\def\ZZ{\mathbb Z}

\def\fg{\mathfrak g}  \def\fp{\mathfrak p}

\def\R{\mathbb{R}}
\def\C{\mathbb{C}}
\def\Z{\mathbb{Z}}
\def\lie#1{\ensuremath{\mathfrak{#1}}}

\DeclareMathOperator{\Sym}{Sym}
\DeclareMathOperator{\Vect}{Vect}
\DeclareMathOperator{\Spin}{Spin}
\DeclareMathOperator{\id}{id}
\def\d{{\rm d}}
\def\bu{\bullet}

\def\so{\lie{so}}
\def\U{\ensuremath{\mathrm{U}}}
\def\SU{\ensuremath{\mathrm{SU}}}

\begin{document}

\allowdisplaybreaks

\newcommand{\arXivNumber}{2506.02246}

\renewcommand{\PaperNumber}{100}

\FirstPageHeading

\ShortArticleName{Higher Symmetries in Twisted Eleven-Dimensional Supergravity}

\ArticleName{Higher Symmetries in Twisted Eleven-Dimensional \\ Supergravity}

\Author{Fabian HAHNER~$^{\rm a}$, Natalie M. PAQUETTE~$^{\rm a}$ and Surya RAGHAVENDRAN~$^{\rm bc}$}

\AuthorNameForHeading{F.~Hahner, N.M.~Paquette and S.~Raghavendran}

\Address{$^{\rm a)}$~Department of Physics, University of Washington, \\
\hphantom{$^{\rm a)}$}~3910 15th Ave NE, Seattle, WA 98195, USA}
\EmailD{\mail{fhahner@uw.edu}, \mail{npaquett@uw.edu}}

\Address{$^{\rm b)}$~Department of Mathematics, Yale University, P.O.~Box 208283, New Haven, CT 06520, USA}
\EmailD{\mail{surya.raghavendran@yale.edu}}

\Address{$^{\rm c)}$~School of Mathematics, University of Edinburgh, Peter Guthrie Tait Road,\\
\hphantom{$^{\rm c)}$}~King's Buildings, Edinburgh, EH9 3FD, UK}

\ArticleDates{Received July 09, 2025, in final form November 25, 2025; Published online December 03, 2025}

\Abstract{In supersymmetric theories, protected quantities can be reorganized into holo\-morphic-topological theories by twisting. Recently, it was observed by Jonsson, Kim and Young that residual super-Poincar\'e symmetries in certain twisted theories can receive higher corrections, turning them into $L_\infty$ algebras with non-strict actions on the twisted fields. In this note, we show that the same phenomenon occurs for the two admissible twists of eleven-dimensional supergravity. Along the way, we discuss in detail the connection between components of physical and twisted fields. }


\Classification{83E50; 17B81; 17B55}

\section{Introduction and review} \label{sec: intro}

For better or worse, we have not yet said all there is to say about supersymmetric field theories. Progress in their study is largely owed to the presence of BPS observables, which are protected and hence exactly computable at weak coupling. \textit{Twisting} offers a modern perspective on such observables, which isolates them and casts them as associated to a more tractable holomorphic-topological theory. Lessons and applications of such a re-organization are manifold. For one, local operators in holomorphic-topological theories enjoy certain secondary operations which are codified by way of tools from higher algebra \cite{CG1, CG2, williamswang}; the twisted perspective tells us that operator products of protected observables in supersymmetric theories are controlled by the very same structures (\cite{SecondaryOps,Budzik:2022mpd, sucotwist, ElliottSafronov, Gaiotto:2024gii, Garner:2023wrc, garnerhiggscoulomb}; see \cite{Garner:2022its} for a pedagogical review). More ambitiously, one may hope that tests of dualities in terms of BPS quantities may be articulated in terms of twists, a program which has successfully been applied to exact checks of the AdS/CFT correspondence \cite{CostelloMtheory1, CostelloMtheory2,Costello:2018zrm, Costello:2020jbh}.

More explicitly, twisting is a fixed-point or localization construction; given a square-zero supercharge, one may consider a smaller theory whose fields are those invariant under the odd commutative subalgebra generated by that supercharge. However, the condition of being invariant must be imposed homologically, as in the BV-BRST formalism. We remind the reader that in the BV-BRST formalism, one enlarges the space of physical fields by ghosts, antifields, etc. and encodes equations of motions and gauge transformations by way of a BV-BRST differential. Dually, one may think of this data as equipping the (shifted) space of fields $\mathcal{E}_{\rm BV}[-1]$ with the structure of an $L_\infty$ algebra whose brackets encode the Taylor components of the BV-BRST differential. We expand a bit on this notion in the following subsection. When we refer to the space of fields of a theory, this is the object we will mean.

In twisting, one first augments the usual BV-BRST differential by a choice of square-zero supercharge. Doing so deforms the $L_\infty$ structure on the space of fields; after possibly shearing the ghost number grading by an $R$-symmetry, this deformation of the original BV-BRST theory can be regarded as a ``cochain level model'' for the twist. Then one typically passes to a smaller quasi-isomorphic (i.e., isomorphic at the level of its cohomology) model of this cochain complex. This smaller model is typically called the twisted theory. Physically, this second step can be viewed as integrating out fields that are rendered nondynamical by the twisting differential. Its effect can be described in terms of the aforementioned deformed $L_\infty$ structure via homotopy transfer.\looseness=-1

In theories of supergravity, the twisting procedure becomes more subtle, since the action of supersymmetry is gauged. Indeed, supergravity in the first order formalism can be viewed as a theory of connections for the super-Poincar\'e group. The fields on flat space in perturbation theory around the trivial connection schematically take the following form
\begin{equation}\label{table:BV sugra fields}
\begin{tikzcd}[column sep=small, row sep=tiny]
& -1 & 0 & \\
+ & \Omega^0 \bigl(\R^d; \mathfrak{so}(d) \bigr) & \Omega^1 \bigl(\R^d; \mathfrak{so}(d) \bigr)_\omega \\
+ & \Omega^0 \bigl(\R^d; \R^d \bigr)_\xi & \Omega^1 \bigl(\R^d; \R^d\bigr)_e \\
- & \Omega^0 \bigl(\R^d; \Pi S \bigr)_q & \Omega^1 \bigl(\R^d; \Pi S\bigr)_\Psi \\
\end{tikzcd} + \text{antifields \ \dots}
\end{equation}

The space of fields carries a $\Z\times \Z/2$ grading by ghost number and fermion parity. Subscripts denote a generic element of the corresponding summand. The fields denoted $\omega$, $e$, $\Psi$ are components of a connection for the super-Poincar\'e group and give the spin connection, vielbein, and gravitino, respectively. The field $q$, in bidegree $(-1, -)$, is a~ghost for local supertranslations. In \cite{CostelloLi}, the authors defined twisted supergravity as supergravity in perturbation theory around a background where $q$ takes a nonzero vacuum expectation value. The authors provided several strands of justification for this definition, including the observation that twisting a~non-gravitational field theory can be implemented by coupling it to such a~twisted supergravity background.

Twisting in general breaks super-Poincar\'e invariance. More explicitly, given a supersymmetric field theory on flat space, with a super-Poincar\'e algebra $\fp$ of symmetries, the twisting procedure deforms this symmetry algebra to a differential super Lie algebra $(\fp,[Q,-])$ where $Q$ is the twisting supercharge. The homotopy transfer procedure equips the cohomology $\fp_Q = H^\bullet(\fp, [Q,-])$ with the structure of an $L_\infty$ algebra. We call $\fp_Q$ the \textit{residual super-Poincar\'e} algebra of the twisted theory.

As with twisting, the interpretation of $\mathfrak{p}_Q$ in theories of supergravity is more subtle. As illustrated in~\eqref{table:BV sugra fields}, in theories of supergravity, the super-Poincar\'e algebra appears in the space of fields as zero modes of ghosts. Accordingly, we see that the residual super-Poincar\'e algebra describes those zero mode ghosts which are preserved by the VEV given to $q$. It is reasonable to expect that twisted supergravity admits a first order description as a theory of connections for $\mathfrak{p}_Q$.\looseness=-1

It was first noticed in~\cite{JKY} that the $L_\infty$ structure on $\mathfrak{p}_Q$ is non-strict in the case of ten-dimensional super Yang--Mills theory. The even part of $\fp_Q$ includes a parabolic subalgebra $\operatorname{Stab}(Q)\subset \mathfrak{so}(d)$ that stabilizes $Q$; the higher brackets crucially involve the complement of the Levi factor inside this parabolic. Moreover, the action of $\fp_Q$ on the twisted fields involves higher brackets. Typically, the Levi factor of the parabolic in $\operatorname{Stab}(Q)$ determines tangential data necessary to globalize a twist; the presence of higher brackets suggests a correction to this picture.

The authors of \cite{JKY} further suggested that a similar phenomenon occurs for twists of eleven-dimensional supergravity. In this note, we examine this suggestion in detail and find non-trivial~$L_\infty$ structures both in the maximal and minimal twists. In more detail
\begin{itemize}\itemsep=0pt
 \item[--] In the maximal twist, the residual super-Poincar\'e algebra features a four-ary bracket. Interestingly, this bracket depends on the $G_2$ three-form in the topological directions.
 \item[--] In the minimal twist, we find both a three-ary and a four-ary operation on the residual super-Poincar\'e algebra.
\end{itemize}
Further, we explain how these $L_\infty$ algebras act on the fields of the twisted supergravity theories. For this purpose, we construct $L_\infty$ maps $\fp_Q \rightsquigarrow \cE^{Q}_{\rm BV}[-1]$, where $ \cE^{Q}_{\rm BV}[-1]$ denotes the $L_\infty$ algebra of fields of twisted supergravity. We do this by first constructing a map to the minimal model~$L_\infty$ algebra of fields and then lifting this map to component fields.
\begin{itemize}\itemsep=0pt
 \item[--] In the maximal twist, we find a module structure with components $\rho^{(1)}$, $\rho^{(3)}$, and $\rho^{(5)}$. Again, these higher components depend on the $G_2$-structure along the topological directions, even when acting on the minimal model in a flat background that is just given by holomorphic functions on $\CC^2$ with the Poisson bracket.
 \item[--] In the minimal twist, the minimal model of the twisted BV fields is a central extension of the exceptional super Lie algebra $E(5|10)$~\cite{RSW}. We construct an $L_\infty$ map from $\fp_Q$ to this minimal model with components $\rho^{(1)}$, $\rho^{(2)}$, and $\rho^{(3)}$.
\end{itemize}
Twisted supersymmetric theories are known for admitting infinite dimensional enhancements of their finite-dimensional symmetry algebras (see, for example,~\cite{Beem:2013sza, Garner:2023wrc, SCA}). In the case of twisted eleven-dimensional supergravity, these enhanced symmetry algebras are divergence-free vector fields on $\C^2$ in the maximal twist, and $E(5|10)$ in the minimal twist~\cite{LSCA, RSW}. These coincide, up to central extension, with the minimal models (see below) of the twisted BV fields so that the $L_\infty$ maps we construct can also be viewed as a compatibility check between the higher $L_\infty$ operations on $\fp_Q$ and the enhancement to the infinite-dimensional symmetry algebra.

One motivation for studying residual super-Poincar\'e symmetries and their relationship with minimal models for twisted theories is to use them to constrain BPS amplitudes in M-theory on flat space. Indeed, as developed in \cite{perturbiner,Scattering-recursion}, minimal models for BV-BRST spaces of fields can be used to determine and derive relations for tree-level scattering. More ambitiously, one might hope to capture a more analytically tractable portion of the BFSS duality, along the lines of twisted holography \cite{ Costello:2018zrm,CostelloLi, Costello:2020jbh}. Evidence that features of the BFSS duality may factor through twists comes from \cite{MaldacenaHerderschee}, where we noticed that the three-point amplitude (reinterpreted as a~supersymmetric index as described by the authors), is compatible with the maximal twist of eleven-dimensional supergravity. The minimal twist, which preserves 1/16th of the original supercharges \cite{RSW}, and enjoys an infinite dimensional exceptional symmetry, is ripe for applications in this vein.

In the remainder of this introduction, we will provide a brief and, hopefully, friendly review of the relevant homotopical algebra concepts that we make use of in this note, with an emphasis on their physical interpretation. In~Section~\ref{sec: twisted-susy}, we provide a brief review of twisting and the elementary Lie group symmetries inherited from the subgroups of the super-Poincar\'e group which stabilize the corresponding twisting supercharges. We also review the two twists of eleven-dimensional supergravity. Along the way, we show how fields in the twisted theory arise from combinations of fields in physical eleven-dimensional supergravity (viewed as a BV-BRST complex), and emphasize some of the subtleties in this mapping. We hope this will be instructive for the physics readers that may (quite rightly) wonder how the geometric objects characterizing the twisted fields get assembled from components of the eleven-dimensional supergravity multiplet. We also do a similar warm-up example using the holomorphic twist of a~free four-dimensional $\mathcal{N}=1$ vector multiplet, for supergravity-averse readers. In~Section~\ref{sec: max}, we study the maximal twist of eleven-dimensional supergravity. In particular, we compute the $L_{\infty}$ extension of the naive Lie symmetries and its action on twisted component fields. In~Section~\ref{sec: min}, we do the same analysis for the minimal twist, and include a derivation that the maximal twist can be viewed as a further twist of the minimal one.

\subsection{Brief review of homotopical notions} \label{sec: background}
Here we give a brief, intuition-based summary of the aspects of homotopical algebra that are important for this paper. Our precise definitions and conventions follow those in many references which we will cite along the way, and which the reader is encouraged to turn to for the equations.

\subsubsection[L\_infty algebras]{$\boldsymbol{ L_{\infty}}$ algebras}
In the BV-BRST formalism, equations of motion and gauge invariance are imposed homologically by equipping functions on the perturbative space of fields with a BV-BRST differential. This information can be dually described as equipping the perturbative shifted space of fields with the structure of an $L_\infty$ algebra. The reader may be familiar with such structures from their appearance in string field theory \cite{Gaberdiel:1997ia,Zwiebach:1992ie}: they feature an infinite set of higher $n$-ary products~$\ell_n$ (a map taking $n$ objects as input and producing one object as output), satisfying a set of generalized Jacobi identities. In brief, an $L_{\infty}$ algebra is a generalization of the notion of Lie algebra, in which the usual Lie bracket does not necessarily satisfy the ordinary Jacobi identity on the nose, but rather only satisfies it up to corrections known as higher homotopies, which must satisfy a (in general infinite) set of compatibility relations; some references with explicit details and formulas include~\cite{Jurco:2020yyu,Jurco:2018sby}.\footnote{$A_{\infty}$ algebras are similar structures, where associativity fails in a fashion rigidly controlled by homotopies. For a review, and a discussion of how to express these higher brackets in terms of the descent formalism of twisted theories, see \cite{Garner:2022its}.}

Two different $L_{\infty}$ algebras are called \textit{quasi-isomorphic} if there is an $L_\infty$ map between them inducing an isomorphism in cohomology, so any two BV-BRST theories that give quasi-isomorphic~$L_{\infty}$ algebras are classically equivalent. (To take a trivial example, one can always add pairs of decoupled ghost fields without changing the cohomology.) Many computations in this field can be simplified when passing to a judiciously chosen quasi-isomorphic model.

In addition to a differential, the BV-BRST formalism equips functions on the space of fields with an additional piece of structure: the BV anti-bracket, which is a Poisson bracket of odd degree. For a Lagrangian theory, the Poisson bracket can dually be described in terms of a~so-called \textit{cyclic} structure on the space of fields -- a~nondegenerate inner product $\langle -, - \rangle$ that is invariant under the $L_\infty$ operations. Note that the BV-BRST differential, being a derivation of functions on the space of fields, can be thought of as a vector field. In a Lagrangian theory, one can ask for a Hamiltonian which generates this vector field under the anti-bracket, and this is exactly the BV action. Accordingly, the action for a Lagrangian theory takes the form
\begin{equation}\label{BV action}
S(\alpha) = \sum_{n\geq 1} \frac{1}{(n+1)!} \big\langle \alpha, \ell_n \big(\alpha^{\otimes n}\big) \big\rangle.
\end{equation}

\subsubsection{Homotopy transfer}
The primary gadget for transferring $L_{\infty}$ structures across different quasi-isomorphic models is called \textit{homotopy transfer}. Note from~\eqref{BV action} that a differential on the space of fields (an $\ell_1$) corresponds to a quadratic term in the BV action, i.e., a kinetic term. Taking cohomology with respect to $\ell_1$ can therefore be thought of as integrating out fields paired together with this kinetic term. Homotopy transfer is a way of keeping track of classical effective interactions that are generated in this process. This procedure accommodates familiar maneuvers such as integrating out massive KK modes \cite{Zeng:2023qqp}.

Explicitly, the homotopy transfer procedure can be used to transfer algebraic structures on one cochain complex $(L,d)$ to another quasi-isomorphic cochain complex $(L' ,d')$ (which can for example be $H^\bullet(L)$ with zero differential) along a deformation retract, i.e., a diagram of the form%
\begin{equation}
\begin{tikzcd}
	\arrow[loop left]{l}{h}(L , d )\arrow[r, shift left, "p"] & (L' , d')\arrow[l, shift left, "i"]
	\end{tikzcd}\label{eq: retract}
\end{equation}
with maps satisfying
$
 p \circ i = \id_{L'} $, $ i \circ p - \id_L = h \circ d + d \circ h$.
Say $(L,d)$ is equipped with higher brackets making it an $L_\infty$ algebra. Then, the homotopy transfer procedure constructs higher brackets on $(L',d')$ turning it into an $L_\infty$ algebra so that the quasi-isomorphisms of cochain complexes specified in the deformation retract~\eqref{eq: retract} become quasi-isomorphisms in the category of $L_\infty$ algebras. The induced brackets on $L'$ can be expressed in terms of formulas for sums over trees, with vertices being labeled by the brackets on $L$ and internal lines labeled by the homotopy $h$~\cite{LodayVallette}.\looseness=-1

\subsubsection{Minimal models}
Different $L_\infty$-equivalent models of the space of fields render different features of a classical field theory manifest. For example, classical equations of motion for a massless quantum field theory are completely determined by tree-level scattering; this is articulated via the well-known perturbiner expansion. As classical equations of motion are encoded via an $L_\infty$ structure, one may ask for a similar codification of tree-level scattering amplitudes and related structures. As discussed in \cite{perturbiner,Scattering-recursion}, the \textit{minimal model}\footnote{This has nothing to do with the 2d RCFTs of the same name.} of an $L_{\infty}$ algebra can be used to encode both the perturbiner expansion and relations between tree-level scattering such as the Berends--Giele recursion relations in Yang--Mills theory which we solve to obtain the Parke--Taylor formula. More explicitly, any $L_\infty$ algebra is quasi-isomorphic to an $L_\infty$ algebra with vanishing differential called the minimal model. The minimal model is unique up to isomorphism; its underlying vector space is simply the cohomology of the original $L_{\infty}$ algebra. Specifying a deformation retract from the original $L_\infty$ algebra to its cohomology, we can compute the brackets on the minimal model by the homotopy transfer procedure as described above.

Starting from a dg Lie algebra, one can always choose a deformation retraction to its cohomology and ask if higher brackets beyond the restriction of the binary bracket appear after homotopy transfer. If this is not the case, there exists an $L_\infty$ quasi-isomorphism between the original dg Lie algebra and the cohomology with binary bracket. In this case, the dg Lie algebra is called \emph{formal}. Crucially, while the form of the higher brackets depends on the choice of deformation retract, the question whether there are non-vanishing higher brackets at all is independent of such choices. In other words, formality is well-defined property of a dg Lie algebra and does not depend on the choice of homotopy data~\cite{LodayVallette}.

\subsubsection[L\_infty modules]{$\boldsymbol{ L_\infty}$ modules}
 We are most interested in symmetries when they act on things. In the BV-BRST formalism, the notion of a symmetry of a perturbative classical field theory is codified by equipping the space of fields $\mathcal{E}_{\rm BV}[-1]$ with the structure of an $L_\infty$ module. A precise definition can be found in~\cite{LadaMarkl}, but morally such a map includes a series of nonlinear corrections which render exact the failure for the linear part of the map to intertwine Lie brackets. Such a homotopical weakening of the notion of a strict Lie map is flexible enough to accommodate settings where symmetries may only close on-shell or up to a gauge transformation. In the main body of the text, we equip minimal models for the $L_\infty$ algebras describing twists of eleven-dimensional supergravity with the structure of an $L_\infty$-module for the corresponding residual super-Poincar\'e algebras.

\section{Twisted supersymmetry} \label{sec: twisted-susy}
In this section, we will provide a brief review of twisting and establish our notation. We will then turn our attention to the two possible twists of eleven-dimensional supergravity, our primary focus in this note. We demonstrate the maps by which fields in the twisted theory~-- that are identified with various geometric objects on spacetime~-- can be viewed as originating from certain field configurations in the physical theory, including a warm-up treatment of the holomorphic twist of a four-dimensional $\mathcal{N}=1$ vector multiplet. When we speak of twisted supergravity, we shall always work perturbatively around the flat background.

\subsection{Twisted field theories and their residual symmetries}
By definition, any supersymmetric field theory comes with an action of a super Lie algebra of symmetries $\fp$. For Poincar\'e invariant field theories, these are super-Poincar\'e algebras enlarging the even affine symmetries of spacetime by odd elements; in some cases, $\fp$ can be enlarged further to a superconformal algebra. We will denote the decomposition of the algebra into even (bosonic) and odd (fermionic) generators, respectively, as
$
 \fp = \fp_+ \oplus \fp_-$.
Further, we denote the super Lie bracket on $\fp$ by $[-,-]$, where it is understood that the bracket is antisymmetric if at least one input is in $\fp_+$ and symmetric if both inputs are in $\fp_-$.

The twisting procedure extracts protected subsectors from a supersymmetric field theory by choosing an odd square-zero element $Q \in \fp$ and taking invariants with respect to the odd abelian subalgebra spanned by $Q$. Such a subsector, \emph{the twisted theory}, governs certain protected quantities such as BPS states and operators. Since $Q$ transforms as a spinor under the Lorentz group, twisting will generically break Poincar\'e invariance,\footnote{For some topological twists, invariance under a copy of $\Spin(d)$ can be restored by choosing a twisting morphism. However, this is not possible for holomorphic or mixed type twists. For a discussion on the existence of twisting morphisms in various dimensions, see~\cite{ElliottSafronov}.} as described in more detail shortly.

In the BV or BRST formalism, the twisted theory is obtained by deforming the BV or BRST differential by the action of the supercharge $Q$ on the theory,
$
 Q_{\rm BV} \mapsto Q_{\rm BV} + Q$.
We briefly remark that, since $Q$ carries ghost number zero and $Q_{\rm BV}$ is of ghost number one, such a~deformation by $Q$ always breaks the integer ghost number grading of the BV theory. In many examples, an integer grading in the twisted theory can be restored by performing a~regrading using $R$-symmetry charges; in other cases the twisted theory is only $\ZZ_2$-graded (see~\cite{CostelloHol} for a~rigorous discussion).

The moduli space of twists is the nilpotence variety,
\[
 Y = \{ Q \in \fp_- \mid [Q,Q] = 0 \}.
\]
These spaces were classified for super-Poincar\'e algebras in~\cite{NV, ElliottSafronov}; in these cases, the nilpotence variety is acted upon by Lorentz transformations and $R$-symmetry and the orbit stratification for this combined action classifies the different twists of a field theory with symmetry algebra $\fp$.

\begin{Remark}
 In theories of supergravity, where supersymmetries are viewed as gauge symmetries, the interpretation of the twisting procedure is subtle, especially when working in non-flat backgrounds. In~\cite{CostelloLi}, these questions were clarified by defining twists of supergravity theories by placing them in backgrounds where the supersymmetry ghosts take a non-zero value. In~a~flat background, this value for the supersymmetry ghost precisely corresponds to the twisting supercharge $Q$.
\end{Remark}

\subsubsection{Residual symmetries}
What are the symmetries of a twisted theory? Clearly, the presence of the twisting supercharge breaks $\fp$ to the subalgebra that preserves $Q$. Further, elements of $\fp$ that are in the image of~${[Q,-]}$ are to be viewed as trivial in the twisted theory. More generally, the choice of a square zero-supercharge $Q$ equips the symmetry algebra $\fp$ with the structure of a differential super Lie algebra
$
 \fp \mapsto (\fp, [Q,-])$,
where the differential is simply $Q$ itself, acting via the bracket.
Its cohomology $\fp_Q := H^\bullet((\fp ,[Q,-]))$ is the residual symmetry algebra of the twisted theory. This means that twisting generally both breaks symmetries (elements which are not in the kernel of $[Q,-]$) and trivializes symmetries (elements which are in the image of $[Q,-]$). The residual symmetry algebra $\fp_Q$ has its own nilpotence variety $Y_{Q}$ encoding the possible further twists of the theory generated by the remaining odd square-zero supercharges.

The super Lie bracket on $\fp$ descends to a bracket on $\fp_Q$, and thereby endows it with the structure of a super Lie algebra. However, viewing the residual symmetries simply as a super Lie algebra is not the full story. Instead, fixing a deformation retract (cf. Section~\ref{sec: background})
\[
	\begin{tikzcd}
	\arrow[loop left]{l}{h}(\fp , [Q,-] )\arrow[r, shift left, "p"] & (\fp_Q , 0)\arrow[l, shift left, "i"]
	\end{tikzcd}
\]
and performing the homotopy transfer can lead to additional higher operations that turn $\fp_Q$ into an $L_\infty$ algebra. The fact that this $L_\infty$ structure is not necessarily strict was first observed in~\cite{JKY} for the holomorphic twist of ten-dimensional $\cN=1$ supersymmetry.
We emphasize that, while the precise form of the higher operations generally depends on the deformation retract, the question whether higher brackets appear at all (i.e., whether $(\fp, [Q,-])$ is formal as a dg Lie algebra) is independent of these choices.

While the above discussion is applicable for any super Lie algebra $\fp$ and twisting supercharges~$\fp_Q$ acting on a field theory, we are in the following, interested in the case where $\fp$ is a super-Poincar\'e algebra. In this case we will call $\fp_Q$ the \emph{residual super-Poincar\'e algebra}.

\subsubsection{Residual super-Poincar\'e algebras}
Let us fix a super-Poincar\'e algebra $\fp$ and discuss the residual symmetries for this case specifically. The even piece of $\fp$ is a semidirect product $(\lie{so}(d) \oplus \fg_R) \ltimes V$ between Lorentz transformations and $R$-symmetries, with the translations living in the vector representation $V$ of $\lie{so}(d)$. This is enlarged by odd elements sitting in a spin representation of $\lie{so}(d)$ and a representation of the $R$-symmetry specified by the choice of one or two integers $\cN$ counting the number of (possibly chiral) supercharges, depending on spacetime dimension. Denoting this representation by $\Sigma$, the bracket between two odd elements is specified by a map
$
 \gamma\colon \Sym^2(\Sigma) \longrightarrow V$.
In particular, the usual $\ZZ_2$-grading that tracks even and odd elements of $\fp$ can be conveniently lifted to a~$\ZZ$-grading by setting $\fp_0 = \lie{so}(d) \oplus \fg_R$, $\fp_1 = \Sigma$, and $\fp_2 = V$. We will work in this lifted $\ZZ$-grading for the rest of this paper.

One can read off many properties of the twisted theory directly from its residual super-Poincar\'e algebra. For example, the twisted theory depends in a generically mixed topological-holomorphic manner on the underlying spacetime. This dependence is encoded by the surviving, nontrivial translation generators which live in
\[
(\fp_Q)_2 = H^2(\fp,[Q,-]) = V/\mathrm{Im}([Q,-]) .
\]
On general grounds, for non-vanishing $Q$, the dimension of $(\fp_Q)_2$ is at most half the dimension of the vector representation~(see, for example,~\cite{ElliottSafronov})
$
\dim((\fp_Q)_2) \leq \frac{1}{2} \dim V $.
Depending on the dimension of this cohomology group, one distinguishes different cases.
\begin{itemize}\itemsep=0pt
	\item[--] If $(\fp_Q)_2=0$, all translations act trivially on the twisted theory. The supercharge $Q$ is called topological.
	\item[--] If the dimension of $V$ is even and the inequality is saturated, precisely half of the translations act trivially on the twisted theory. Choosing such a supercharge induces a complex structure on spacetime; the supercharge $Q$ is called holomorphic.
	\item[--] In the generic case, when more than half, but not all, translations act trivially, the supercharge $Q$ is called mixed.
\end{itemize}
In general, twisting a field theory on $\R^d$ by a supercharge with
$
\dim((\fp_Q)_2) = k
$
yields a topologic-holomorphic theory formulated on $\RR^{d-2k} \times \CC^k$. We say that $Q$ has $d-2k$ topological directions (and $2k$ holomorphic directions).

The degree one piece $(\fp_Q)_1 = H^{1}(\fp , [Q,-])$ contains those odd supercharges that can act non-trivially on the twisted theory. Note that for $(\fp_Q)_1$ both the kernel and the image condition are generally non-trivial, meaning that some of the original supersymmetries in $\fp$ are broken (those not commuting with $Q$) while others are trivialized (those in the image of $[Q,-])$. Square-zero elements in $(\fp_Q)_1$ define the nilpotence variety $Y_{Q}$ which encodes further twists that can be obtained from the $Q$-twisted theory.

The nilpotence variety $Y_{Q}$ is acted upon by the Lie group exponentiating residual super-Poincar\'e transformations in degree zero $(\fp_Q)_0 = H^0(\fp, [Q,-])$ which is the commutant of $Q$. These are precisely those Lorentz transformations and $R$-symmetries preserving the twisting supercharge $Q$.

\subsubsection{Symmetry enhancements}
One interesting feature of twisted field theories is that they typically feature an enhancement of the finite-dimensional residual symmetries $\fp_Q$ to an infinite-dimensional (super) Lie algebra. This was first observed in the context of superconformal field theories in four-dimensions where the Virasoro algebra appears for certain twists by supercharges of the superconformal algebra~\cite{Beem:2013sza}. It was later realized that this is a more generic features of twisted theories with at least some holomorphic directions (see, for example,~\cite{Bomans:2023mkd, Garner:2023wrc, RSW, RW,SCA}). Typically these infinite-dimensional symmetry algebras involve holomorphic vector fields along the subspace of spacetime on which the twisted theory behaves holomorphically. A~systematic treatment inspired by the pure spinor superfield formalism was developed in~\cite{LSCA}.

Denoting the enhanced infinite-dimensional symmetry algebra by $\cG$, compatibility requires that there is an $L_\infty$ map
$\fp_Q \rightsquigarrow \cG $.
In the cases of twisted eleven-dimensional supergravity, where $\cG$ coincides~-- up to a central extension~-- with the minimal model of the BV fields itself, we construct these maps in Sections~\ref{sec: max-module} and~\ref{sec: min-module}.

\subsubsection{Twists and curved backgrounds}
The twisting procedure, as we introduced it, requires the existence of a global square-zero twisting supercharge. As most backgrounds break supersymmetry, this comes with severe restrictions, e.g., on the holonomy group of spacetime. More sophisticated procedures to preserve supersymmetry in curved backgrounds involve non-minimal couplings to supergravity~\cite{Festuccia:2011ws}.\footnote{For a discussion of global aspects of twists specifically in four dimensions see also~\cite{Cushing:2023rha}.} On the other hand, twisting a supersymmetric field theory in a flat background $\R^d$, we obtain a topological-holomorphic field theory on $\R^{d-2k} \times \C^k$. Such theories are well defined on any manifold $M$ equipped with a transversely holomorphic foliation of type $(d-2k, k)$. These are precisely those manifolds that are equipped with a compatible atlas whose charts map open neighborhoods of~$M$ to $\R^{d-2k} \times \C^k$. In other words, $M$ is equipped with coordinates $(x_1, \dots, x_{d-2k}, z_{1}, \dots , z_k)$ and coordinate changes are of the form $(f(x,z), g(z))$, where $g$ is a holomorphic function. In this work, we study higher structures arising in the residual symmetries of twisted theories in flat backgrounds, but we expect similar structures to also appear in more complicated backgrounds. These symmetries could further illuminate the geometric structures amenable to the twisting procedure. We hope to extend this line of thought in future work.

\subsection{Eleven-dimensional supergravity and its twists}
Let $V$ denote the vector representation of $\Spin(11)$ and $S$ be the unique spinor representation of dimension 32. The super-Poincar\'e algebra\footnote{Here and in the following, we work with the complexified version of the super-Poincar\'e algebra. This is the natural setting for twisted supergravity; indeed, the existence of twisting supercharges relies on complexification~\cite{NV,ElliottSafronov}.} is then given by
$
 \fp_0 = \so(V)$, $\fp_1 = S$, $\fp_2 = V $.
Eleven-dimensional supergravity is the low energy limit of M-theory. Its component fields were first described in~\cite{CJS}; the physical fields are the graviton $g_{\mu \nu}$, the gravitino $\psi_\mu^\alpha$, and a three-form gauge field \smash{$C^{(3)}$}. In the BV formalism, the theory is modeled by a local cyclic $L_\infty$ algebra built from these fields together with their ghost systems and antifields. In more detail, there are diffeomorphism ghosts $\xi_\mu$ which are vector fields on the spacetime thought of as infinitesimal diffeomorphisms, as well as a bosonic ghost $q$ for the supercharges taking values in $S$; further, the three-form carries a ghost system consisting of two-, one-, and zero-forms that we denote by $C^{(i)}$ for $i=0,1,2$. In addition each of these fields and ghosts comes with its corresponding antifield denoted by a superscript $^+$.

The underlying cochain complex of this $L_\infty$ algebra is described in the following diagram, where $\Omega^{k}$ denote differential forms of degree-$k$ and $T$ denotes sections of the tangent bundle,
\begin{gather}
\Omega^{0} \xrightarrow{\d}
\Omega^{1} \xrightarrow{\d}
\Omega^{2} \xrightarrow{\d}
\Omega^{3} \xrightarrow{\d \star \d}
\Omega^{8} \xrightarrow{\d}
\Omega^{9} \xrightarrow{\d}
\Omega^{10} \xrightarrow{\d}
\Omega^{11},\nonumber
\\
T \xrightarrow{\mathcal{L}_\xi g}
\mathrm{S}^2(T) \xrightarrow{E}
\mathrm{S}^2(T) \xrightarrow{\cL^*}
T,\qquad
S \otimes \Omega^0 \xrightarrow{\d}
S \otimes \Omega^1 \xrightarrow{D_{\rm RS}}
S \otimes \Omega^{10} \xrightarrow{\d}
S \otimes \Omega^{11}.\label{eq: sugrafields}
\end{gather}
Here, the horizontal axis is ghost number.
As usual, the differential encodes the linearized part of the equations of motion and gauge invariances. For the first row describing the three-form, gauge transformations are simply given by the de Rham differential; in terms of local operators we can write
\smash{$
 \delta C^{(3)} = \d C^{(2)}$},
and similarly for the one- and zero-form ghosts that encode dependencies between gauge transformations. As indicated by the differential, the linearized equations of motion for the three-form are simply $\d \star \d C^{(3)}=0$.
The second row describes the graviton as a symmetric two-tensor that perturbs the background metric. The diffeomorphism ghosts are vector fields that act on the graviton by the Lie derivative along the background metric. The differential acting on the graviton field is given by the linearized Einstein equation (here indicated by the map $E$).
Finally, the third row contains the gravitino subject to the Rarita--Schwinger equation and gauged by the supersymmetry ghost.

The higher brackets of the $L_\infty$ structure describe the interactions and the non-linear contributions to the gauge transformations. For example, there is a binary bracket for the three-form field,
\begin{equation}
 \mu_2\colon\ \Omega^3 \otimes \Omega^3 \longrightarrow \Omega^8, \qquad \mu_2\big( C^{(3)}, C^{(3)}\big) = \d C^{(3)} \wedge \d C^{(3)}.
\end{equation}
Using the cyclic pairing on~\eqref{eq: sugrafields} that pairs fields and antifields (e.g., $\Omega^{11}$ with $\Omega^0$ etc.), this binary bracket corresponds to the Chern--Simons interaction term in the action functional, here arising via $\big\langle C^{(3)} , \mu_2\big(C^{(3)}, C^{(3)}\big) \big\rangle = \int C^{(3)} \wedge \d C^{(3)} \wedge \d C^{(3)} $. Note that the interactions for the graviton give rise to an infinite series of $L_\infty$ operations $\mu_k$ that correspond to the expansion of the Einstein--Hilbert action around a flat background.

\subsubsection{The supersymmetry module structure}
The BV fields form an $L_\infty$ module for the super-Poincar\'e algebra $\fp$. The linear part of the module structure is given by the obvious action of the Poincar\'e algebra on the fields together with the supersymmetry transformation described in~\cite{CJS}. However, working with the BV complex of component fields presented above, the action of supersymmetry only closes on-shell on the component fields, which means that the BV fields do not form a Lie module. Instead, there are additional higher terms in the action making it an $L_\infty$ action. The full $L_\infty$ action is described in a language similar to that of this note in~\cite{MaxTwist} (see~\cite{BerkovitsSupermembrane} for an earlier account).\footnote{Stritifications of this $L_\infty$ module structure exist in the context of the pure spinor superfield formalism (see \cite{Ced-11d, CY2}). From the perspective of the bosonic spacetime, this amounts to the introduction of an infinite set of auxiliary fields.}

In theories of supergravity, supersymmetry becomes a gauge symmetry. Thus, we can recover the action of $\fp$ by restricting the ghost fields to certain values; in other words, the action of~$\fp$ on the BV fields is inner.
This means it can be described by an $L_\infty$ map into the BV fields\footnote{We also use the notation for cohomological degree shifts such that for a graded complex $V = \oplus_j V_j$, the shift~$V[i]$ maps the graded subspaces via $V[i]_j = V_{i + j}$. For example, if $V = V_0$ is a space of ordinary physical fields concentrated in cohomological degree zero, then $V[-1]$ is concentrated in degree 1.}~${
 \rho \colon \fp \rightsquigarrow \mathcal{E}_{\rm BV} [-1]}$,
where $\mathcal{E}_{\rm BV}$ denotes the space of fields in the BV formalism.
Explicitly, the linear component of~$\rho$ maps $\so(V) \ltimes V$ to vector fields specifying values for the diffeomorphism ghost
\[
 \rho^{(1)}(v) = v^\mu \frac{\partial}{\partial x^\mu}, \qquad v \in V ,\qquad
 \rho^{(1)}(M) = M^{\mu \nu} x_\mu \frac{\partial}{\partial x^\nu} ,\qquad M \in \so(V).
\]
These vector fields generate translations and rotations in spacetime respectively; the diffeomorphism ghost acts on the fields in the BV complex by the Lie derivative thereby reproducing the standard action of the Poincar\'e algebra on the fields of the supergravity theory. Similarly, the odd elements of $\fp$ are mapped to covariantly constant configurations of the supersymmetry ghosts, reproducing the supersymmetry transformations of the fields.

\subsubsection{The nilpotence variety and twists}
Investigating the nilpotence variety, one finds two distinct twists available to the eleven-dimen\-sion\-al supergravity theory~\cite{NV}:
\begin{itemize}\itemsep=0pt
 \item[--] The maximal twist is topological in seven directions and holomorphic in the remaining four directions. Starting from a flat background, it thus places the theory in $\RR^7 \times \CC^2$.
 \item[--] The minimal twist is holomorphic in ten directions and topological in the remaining one. Thus, it identifies the flat background with $\C^5 \times \RR$.
\end{itemize}

\subsubsection{A first look at the maximal twist}\label{p: first-max}
The maximal twist of eleven-dimensional supergravity in a flat background is Poisson--Chern--Simons theory on $\CC^2 \times \RR^7$. This description was first suggested by Costello in~\cite{CostelloMtheory2}. The conjecture was verified at the free level in~\cite{MaxTwist} and realized as a further twist of minimally twisted supergravity in~\cite{RSW}. In~\cite{CY2}, pure spinor superfield techniques are used to provide a~computation of the twist at the interacting level.

Poisson--Chern--Simons theory is described by the local $L_\infty$ algebra
\[
 \big( \Omega^{(0,\bullet)}\big(\CC^2\big) \otimes \Omega^\bullet\big(\RR^7\big) , \bar{\partial}_{\CC^2} + \d_{\RR^7} , \{-,-\} \big).
\]
Here, the binary operation is the Poisson bracket; choosing coordinates $(z_1,z_2)$ on $\CC^2$, it can be expressed in terms of the Poisson-bivector $\pi= \partial_{z_1} \wedge \partial_{z_2}$,
$
 \{\alpha,\beta\} = \pi(\partial \alpha \wedge \partial \beta) $.
Poisson--Chern--Simons theory can be formulated on any product $X\times M$ between a Calabi--Yau twofold $X$ and an odd-dimensional real manifold $M$. Denoting the holomorphic volume form on $X$ by $\Omega$, the BV action of the theory reads
\[
	S_{\rm BV}(\alpha) = \int_{X \times M} \Omega \wedge \alpha \left( \frac{1}{2}\big(\bar{\partial}_X + \d_M\big) \alpha + \frac{1}{6} \{ \alpha , \alpha \} \right) .
\]
We remark that the theory is generally only $\ZZ_2$-graded (i.e., there is no well defined integer ghost number, so that there is no invariant distinction between the ghosts, antifields, and physical fields in the twist). Only if the real manifold $M$ is one-dimensional, the $\ZZ$-grading can be restored for degree reasons.\footnote{Recall that a $\ZZ$-graded BV theory is equipped with a $(-1)$-shifted symplectic structure on the BV fields $\cE_{\rm BV}$ inducing a Poisson bracket of degree $+1$ on the observables (often called the antibracket). In terms of the $L_\infty$ algebra $\cE_{\rm BV}[-1]$ this structure corresponds to a cyclic pairing of degree $-3$. In a $\ZZ_2$-graded BV theory one only has an \emph{odd} shifted symplectic structure, or equivalently, an odd pairing on the $L_\infty$ algebra.}

\subsubsection{A first look at the minimal twist}
The component fields of minimally twisted eleven-dimensional supergravity at free level were described in~\cite{Ced-SL5,spinortwist}, interactions were conjectured (and numerous consistency checks performed) in~\cite{RSW}; a geometric pure spinor construction of the interaction was given in~\cite{CY2}.

Explicitly, the free component fields on $\CC^5 \times \RR$ are the $\ZZ/2\ZZ$ graded BV theory described by the following cochain complex:
\begin{equation} \label{eq: comp-min-tw}
\begin{bmatrix}
\big(\Omega^{0,\bu}\big(\CC^5\big) ,\bar{\partial}\big)_\beta \xrightarrow{\partial} \big(\Omega^{1,\bu}\big(\CC^5\big) , \bar{\partial}\big)_\gamma \\
\big(\mathrm{PV}^{1,\bu}\big(\CC^5\big) , \bar{\partial}\big)_\mu \xrightarrow{\partial_{\Omega}} \big(\mathrm{PV}^{0,\bu}\big(\CC^5\big),\bar{\partial}\big)_\nu \\
\end{bmatrix} \otimes \big(\Omega^\bu(\R) ,\d\big).
\end{equation}
Here, $\mathrm{PV}^{k,\bu}\big(\CC^5\big) = \Omega^{0,\bu}\big(\CC^5 , \wedge^k T_{\CC^5}^{(1,0)}\big)$ denotes polyvector fields understood as Dolbeault forms with values in exterior powers of the holomorphic tangent bundle and $\partial_\Omega$ denotes the divergence operator associated to the standard holomorphic volume form $\Omega = \d z_1 \wedge \dots \wedge \d z_5$. The subscript denotes notation for a generic element of the indicated summand tensored with de~Rham forms on $\R$. This cochain complex can be equipped with an $L_\infty$ structure turning it into an interacting BV theory. The action for this BV theory is a deformation of one of generalized BF-type \cite{RSW}. Explicitly, the interaction of the theory takes the form
\[
\int \Omega \wedge \frac{\partial\gamma \vee \mu^2}{1-\nu} + \int \gamma\partial\gamma\partial\gamma.
\]

The first term above is interpreted as a geometric series in $\nu$ and can be related to a BF-type extension of the BCOV interaction by way of a nonlinear field redefinition. The second term is reminiscent of the Chern--Simons term of eleven-dimensional supergravity.

In this paper we will also work with a theory of field strengths for the minimal twist of eleven-dimensional supergravity. The component fields read
\begin{equation} \label{eq: comp-min-tw-str}
\begin{bmatrix}
\big(\Omega^{2,\bu}\big(\CC^5\big) , \bar{\partial}\big) \xrightarrow{\partial} \big(\Omega^{3,\bu}\big(\CC^5\big) , \bar{\partial}\big) \xrightarrow{\partial} \big(\Omega^{4,\bu}\big(\CC^5\big) , \bar{\partial}\big) \xrightarrow{\partial} \big(\Omega^{5,\bu}\big(\CC^5\big) , \bar{\partial}\big) \\
\big(\mathrm{PV}^{1,\bu}\big(\CC^5\big) , \bar{\partial}\big) \xrightarrow{\partial_{\Omega}}\big(\mathrm{PV}^{0,\bu}\big(\CC^5\big),\bar{\partial}\big)
\end{bmatrix} \otimes \big(\Omega^\bu(\R) ,\d\big).
\end{equation}

We claim that we can equip this cochain complex with the structure of a degenerate BV theory~-- i.e., it admits a degenerate copairing instead of a nondegenerate pairing. We will not need the full structure of the BV theory here so we leave explicating it to elsewhere. We instead highlight some essential features:
\begin{itemize}\itemsep=0pt
 \item[--] The strict map that acts as $\partial$ on $\gamma$, $0$ on $\beta$, and the identity on $\mu$ and $\nu$ defines a map of BV theories from the minimal twist of eleven-dimensional supergravity to the theory of field strengths.
 \item[--] The minimal model on flat space is $L_\infty$ equivalent to $E(5|10)$, with no central extension.
\end{itemize}

\subsection{How is the twisted theory related to the physical fields?}
In the BV formalism, twisted theories are described by local $L_\infty$ algebras that typically involve de Rham forms along the topological directions and $(0,\bullet)$-forms along the holomorphic direction, potentially with values in a graded vector bundle. While this abstract description makes the topological-holomorphic nature of the twisted theory explicit, it raises the obvious question how the degrees of freedom in the twisted theory precisely arise in terms of the component fields of the full physical theory. In other words: how is the protected sector that the twisted theory describes precisely related to the full theory?

\subsubsection{An instructive example: the vector multiplet in four dimensions}
Let us consider the holomorphic twist of a single vector multiplet for four-dimensional $\cN=1$ supersymmetry. Recall that we have the exceptional isomorphism $\Spin(4) \cong \SU(2)_+ \times \SU(2)_-$; the two spin representations $S_\pm$ are the fundamental representations of the first and the second~$\SU(2)$ respectively. We denote the vector representation by $V$. Then the $\cN=1$ super-Poincar\'e algebra is of the form
$
 \fp_0 = \so(4)$, $\fp_1 = S_+ \oplus S_- $, $\fp_2 = V$.
The bracket between two odd elements is given by the identification $S_+ \otimes S_- \cong V$. Square-zero elements are precisely those that are either completely in $S_+$ or in $S_-$. This means that the nilpotence variety $Y=S_+ \cup_{\{0\}} S_-$ is the union of two planes intersecting in the origin. All points in either plane are related by the action of $\Spin(4)$ so that $Y$ has two non-trivial orbits representing the two possible twists; both twists are holomorphic.

The BRST fields of the abelian vector multiplet are described by the following cochain complex
\[
 \begin{tikzcd}[row sep=tiny]
 \Omega^0 \arrow[r,"\d"] & \Omega^1 \\
 & S_+ \oplus S_- \\
 & \Omega^0.
 \end{tikzcd}
\]
The field content is a one-form gauge field $A \in \Omega^1$ together with its ghost field $c \in \Omega^0$, as well as the gaugino $\big(\lambda, \bar{\lambda}\big) \in S_+ \oplus S_-$ and an auxiliary field $D \in \Omega^0$.

Let us choose a twisting supercharge $Q \in S_+$. It is well known that the twist of the BRST multiplet is given by $\big(\Omega^{(0,\bu)}\big(\CC^2\big) , \bar{\partial}\big)$~(\cite{JohansenHet, nikitathesis}; see, for example,~\cite{SWchar} for a discussion in a language similar to this note, and \cite{Budzik:2023xbr,Budzik:2022mpd} for more). In order to make this identification, one decomposes the field content, supersymmetry transformations, and BRST differential under $\mathrm{SU}(2)_- \times \U(1)$ where the $U(1)$ is the Cartan of $\mathrm{SU}(2)_+$. This means that the vector and spinor representations decompose as follows:
$ S_+ \rightarrow \textbf{1}^{+1} \oplus \textbf{1}^{-1}$, $ S_- \rightarrow \textbf{2}^0$, $V \rightarrow \textbf{2}^1 \oplus \textbf{2}^{-1}$.
After applying a~twisting morphism mixing the $\U(1)$ charge with the $R$-symmetry charge (see~\cite{SWchar} for details on the gradings) we decompose the field content. Under the new grading the twisting supercharge has~$\U(1)$ weight zero; further we can identify the grading with a weight on Dolbeault forms by putting~${|\d z| =1}$ and $|\d \bar{z}|=-1$. For the one-form gauge field and its ghost, we identify~${
 c \in \Omega^{(0,0)}} $ and $ \big(A, \bar{A}\big) \in \Omega^{(1,0)} \oplus \Omega^{(0,1)}$.
The gaugino and the auxiliary field take values in the representations
\[
 (\lambda_+ , \lambda_-) \in \textbf{1}^{0} \oplus \textbf{1}^{-2}, \qquad \bar{\lambda} \in \textbf{2}^1, \qquad D \in \textbf{1}^0.
 \]
It is instructive to identify these with subrepresentations of $\Omega^{\bu,\bu}\big(\CC^2\big)$; $\bar{\lambda}$ is identified with a~form in \smash{$\Omega^{(1,0)}$}, $\lambda_-$ becomes a form in~\smash{$\Omega^{(0,2)}$}, while $\lambda_+$ and $D$ can be viewed as $(1,1)$-forms in the subspace spanned by the antisymmetrized representative~${\d z_1 \wedge \d \bar{z}_2 - \d z_2 \wedge \d \bar{z}_1}$. We denote this subspace by \smash{$\Omega^{(1,1)}_{\textbf{1}}$}. With these identifications, the twisted BRST multiplet decomposes as follows:
\[
 \begin{tikzcd}
 \Omega^{(0,0)}_c \arrow[d, "\textstyle\partial"] \arrow[r, "\textstyle \bar{\partial}"] & \Omega^{(0,1)}_{\bar{A}} \arrow[r, "\textstyle\bar{\partial}"] \arrow[d, "\textstyle\sigma \circ \partial"] & \Omega^{(0,2)}_{\lambda_-} \\
 \Omega^{(1,0)}_A \arrow[r, "\textstyle\sigma \circ \bar{\partial}"] & \big(\Omega^{(1,1)}_\textbf{1}\big)_{\lambda_+} \\
 \Omega^{(1,0)}_{\bar{\lambda}} \arrow[u, "\cong"] \arrow[r, "\textstyle \sigma \circ \bar{\partial}"] & \big(\Omega^{(1,1)}_\textbf{1}\big)_{D} \arrow[u, "\cong"].
 \end{tikzcd}
\]
Here, $\sigma$ denotes the map projecting $(1,1)$-forms to $\Omega^{(1,1)}_\textbf{1}$. The subscripts indicate the origin of these representations in terms of the BRST fields.

Clearly, the first horizontal row forms a copy of $(0,\bullet)$-forms that appear in the twisted theory. To obtain the twisted theory, we take cohomology with respect to those pieces of the differential denoted vertically in the picture. The horizontal $\bar{\partial}$-differential descends to the differential in the twisted theory.

Indeed, we see that there are supersymmetry transformations relating $A$ and $\bar{\lambda}$ as well as~$\lambda_+$ and $D$ so that they form acyclic pairs, which cancel when we pass to cohomology. We can imagine first taking cohomology with respect to these pieces of the differential and see that only the first horizontal row remains.\footnote{This can be made precise by equipping the complex with an appropriate filtration and considering the associated spectral sequence.}

However, the elements sitting in this first horizontal row are not the full cohomology classes. Ignoring the horizontal $\bar{\partial}$-differential for a minute, we find that the complex decomposes into three pieces
\[
 \begin{tikzcd}
 \Omega^{(0,0)}_c \arrow[d, "\partial"] & & \Omega^{(0,1)}_{\bar{A}} \arrow[d, "\sigma \circ \partial"] & & \Omega^{(0,2)}_{\lambda_-}. \\
 \Omega^{(1,0)}_A & \Omega^{(1,0)}_{\bar{\lambda}} \arrow[l, "\cong"], & \big(\Omega^{(1,1)}_\textbf{1}\big)_{\lambda_+} & (\Omega^{(1,1)}_\textbf{1})_{D} \arrow[l,"\cong"],
 \end{tikzcd}
\]
For example, we can identify the cohomology of the first piece with $\Omega^{(0,0)}$, but the appropriate representative of a corresponding cohomology class in terms of component fields is given by a pair $\big(c, \bar{\lambda}\big)$ satisfying the condition that $\partial c = \bar{\lambda}$. This is a general feature of such cochain complexes when presented in terms of ``stairs'' as above: the full cohomology class is a linear combination of pieces sitting on the diagonal of the stair. These contributions can be obtained by ``walking down the stair''; in this case this means first applying $\partial$ and then inverting the isomorphism on $\Omega^{(1,0)}$. (For a mathematical treatment of these techniques see, e.g.,~\cite{Stelzig}; for a~discussion in a physics context see, e.g.,~\cite{BPS-SS}.)

We can use this to write the quasi-isomorphism identifying the forms of the Dolbeault complex appearing in the twisted theory as field configurations in the full theory
$
 i \colon \Omega^{(0,\bullet)}\big(\CC^2\big) \longrightarrow (\cE_{\rm BRST} , Q_{\rm BRST} + Q)$.
Explicitly, on $(0,0)$-forms we have
$ i(c) = c + \partial c$.
For $(0,1)$-forms, we have~${
 i\big(\bar{A}\big) = \bar{A} + \sigma \circ \partial \bar{A}}$,
while the $(0,2)$-form representative does not get corrected at all
$ {i(\lambda_-) = \lambda_-}$.

\begin{Remark}
 Although we considered the BRST complex in this example, the situation does not change much when taking into account the equations of motion. In fact, take the representative of the $(0,0)$ form: its derivative $\partial c$ is viewed as a configuration of the gaugino in the physical theory. The Dirac operator acts on $\bar{\lambda}$ by $\varepsilon^{\alpha \beta} \partial_\alpha \bar{\lambda}_\beta$ so that we immediately find that field configurations~${\bar{\lambda} = \partial c}$ are automatically on-shell.
\end{Remark}

\subsubsection[Maximally twisted eleven-dimensional supergravity in terms of the physical fields]{Maximally twisted eleven-dimensional supergravity\\ in terms of the physical fields}\label{p: max-stair}

We can run a similar procedure to identify the representatives of the twisted supergravity multiplet in terms of the physical fields. Since the supersymmetry transformations only close on-shell, the underlying structure of the cochain complex is more complicated and the ``stairs'' contributing to the embedding have more than just a single step, making the full form of the representatives much more complicated.

Recall that the field content in the twisted theory organizes is $\Omega^{(0,\bullet)}\big(\CC^2\big) \otimes \Omega^\bullet\big(\RR^7\big)$. In~\cite{MaxTwist}, the maximal twist was computed from the component fields by identifying the trivial pairs arising from the action of the twisting supercharge $Q$. This identifies contribution to the representatives from the ``top of the stair'' in terms of the physical fields. For the maximal twist, these are contributions from the three-form with its ghost system and antifields together with certain components of the gravitino and its antifield as summarized in the following table:
\begin{table}[htp]
	\small
	\begin{center}
		\begin{tabular}{c|cccccccc}
			& $\Omega^0\big(\RR^7\big)$ & $\Omega^1\big(\RR^7\big)$ & $\Omega^2\big(\RR^7\big)$ & $\Omega^3\big(\RR^7\big)$ &
			$\Omega^4\big(\RR^7\big)$ & $\Omega^5\big(\RR^7\big)$ & $\Omega^6\big(\RR^7\big)$ & $\Omega^7\big(\RR^7\big)$ \\
			\hline
			\\[-0.3cm]
			$\Omega^{0,0}\big(\CC^2\big)$ & $C^{(0)}$ & $C^{(1)}$ & $C^{(2)}$ & $C^{(3)}$ & $\psi$ &
			$\psi^{+}$ & $C^{(3)+}$ & $C^{(2)+}$ \\ \\[-0.3cm]
			$\Omega^{0,1}\big(\CC^2\big)$ & $C^{(1)}$ & $C^{(2)}$ & $C^{(3)}$ & $\psi$ &
			$\psi^{+}$ & $C^{(3)+}$ & $C^{(2)+}$ & $C^{(1)+}$ \\ \\[-0.3cm]
			$\Omega^{0,2}\big(\CC^2\big)$ & $C^{(2)}$ & $C^{(3)}$ & $\psi$ & $\psi^{+}$ &
			$C^{(3)+}$ & $C^{(2)+}$ & $C^{(1)+}$ & $C^{(0)+}$ \\
		\end{tabular}
	\end{center}
	\caption{The relation between fields in the twisted theory and in the physical theory at the top of the stair.} \label{tab:twistedfields}
\end{table}

One can then use the decomposition of the field content and supersymmetry transformations under $G_2 \times \SU(2)$ to identify the stairs contributing to the representatives in terms of the fields of the full theory. In the following, let us sketch this for the zero-form $\Omega^{(0,0)}\big(\CC^2\big) \otimes \Omega^0\big(\RR^7\big)$ of the twisted theory. The corresponding ``top of the stair'' is simply the zero-form ghost of the three-form ghost system. The stair then takes the following form:
\begin{equation} \label{eq: stair}
 \begin{tikzcd}[ampersand replacement = \&]
 \big(\Omega^{(0,0)} \otimes \Omega^0\big)_{C^{(0)}} \arrow[d, "\partial"] \\
 \big(\Omega^{(1,0)} \otimes \Omega^0\big)_{C^{(1)}} \& T^{(1,0)} \arrow[l, "i_\xi \Omega"] \arrow[d, "\cL_\xi g"] \\
 \& \begin{split}
 T^{(1,0)} \otimes T^{(0,1)} \\ T_{\RR^7} \otimes T^{(1,0)} \\ S^2_0\big(T^{(0,1)}\big)
 \end{split} \& \begin{split}
 & \textbf{3}^{1} \\ &T_{\RR^7} \otimes \textbf{2}^{1} \\ &\textbf{3}^{-1}
 \end{split} \arrow[d, "D_{\rm RS}"] \arrow[l] \arrow[l, shift left = 7.5] \arrow[l, shift right = 7.5] \\
 \& \& \cdots \& \cdots \arrow[l]. 
 \end{tikzcd}
\end{equation}
This means that the zero-form component of the twisted theory gets mapped to a field configuration of the full theory $\big(C^{(0)}, \xi, \psi, \dots\big)$ satisfying induced constraints that can be read off from the diagram. First, we find that $\partial C^{(0)}$ has to be canceled by a supersymmetry transformation of the diffeomorphism ghost
$
 \partial C^{(0)} = i_\xi \Omega$,
where $\Omega = \d z_1 \wedge \d z_2$ is the holomorphic volume form on $\CC^2$. However, the diffeomorphism ghost is mapped by the BRST differential to a field configuration of the graviton. This is given by the Lie derivative along the flat background metric. Recalling that in our coordinates, the metric takes the form
$
 g = \delta_{ab} \d x^a \otimes \d x^b + \delta_{ij} \d z^i \otimes \d \bar{z}^j$,
a short calculation shows that for a vector field of the form $X(x,z,\bar{z})^i \frac{\partial}{\partial z^i}$, the Lie derivative of the metric has three non-vanishing components,
\[
 (\cL_\xi g)_{z_i \bar{z}_j} \in T^{(1,0)} \otimes T^{(0,1)} ,\qquad (\cL_\xi g)_{\bar{z}_i a} \in T_{\RR^7} \otimes \textbf{2}^1, \qquad (\cL_\xi g)_{\bar{z}_i \bar{z}_j} \in \mathrm{S}^2\big(T^{(0,1)}\big).
\]
These contributions are, in turn, canceled by a supersymmetry transformation of the gravitino. However, the stair continues: these gravitino components are mapped to components of the gravitino antifield by the Rarita--Schwinger operator, which are again canceled by supersymmetry transformations of some components of the antifields of other physical fields of the theory etc. We refrain from working this out in full detail here, but emphasize that the fields of the twisted theory are best viewed as some complicated field configuration in the physical theory, where the values of the different ghosts, physical fields and antifields are generated by various derivatives of the field in the twisted theory as determined by such stairs as described above.

\section{Residual symmetries in the maximal twist} \label{sec: max}

Let us now investigate the residual super-Poincar\'e algebra $\fp_Q$ for maximally twisted eleven-dimensional supergravity. In~Section~\ref{sec: max-ht}, we will see how homotopy transfer gives rise to higher operations on $\fp_Q$ making it an $L_\infty$ algebra. Then, we move on to study the action of this $L_\infty$ algebra both on its minimal model (see~Section~\ref{sec: background}) and on the component fields of the twisted theory.

\subsection{The residual super-Poincar\'e algebra and homotopy transfer} \label{sec: max-ht}
Let us fix a maximal twisting supercharge $Q \in \fp_1$. In order to investigate $\fp_Q$, we decompose the super-Poincar\'e algebra under $G_2 \times \SU(2) \times \U(1)$. For this purpose, we use the following inclusion of groups:
\[
 \Spin(11) \supset \Spin(7) \times \Spin(4) \supset G_2 \times \SU(2) \times \U(1).
\]
Here, in the last step we identified $\Spin(4) = \SU(2)_+ \times \SU(2)_-$, and the $\U(1)$ appearing in the last step is the Cartan of $\SU(2)_-$.

The spin representation $S$ of $\Spin(11)$ decomposes as
\[
 S \cong (\textbf{1}_{G_2} \oplus V_7) \otimes \big(\textbf{2}^0 \oplus \textbf{1}^1 \oplus \textbf{1}^{-1}\big).
\]
Our maximal twisting supercharge lives in the summand $Q \in \textbf{1}_{G_2} \otimes \textbf{1}^{-1}$.

Fixing a maximal twisting supercharge determines a decomposition of the vector representation into topological, holomorphic, and anti-holomorphic directions. We will denote this splitting~by
\[
 V = V_7 \oplus L \oplus L^\vee \cong V_7 \oplus \textbf{2}^1 \oplus \textbf{2}^{-1}.
\]
Here, we identified the seven topological directions with the seven-dimensional representation of~$G_2$; holomorphic and anti-holomorphic directions are $L \cong \textbf{2}^1$ and $L^\vee \cong \textbf{2}^{-1}$, respectively.

Remembering\footnote{Here we use the notation $\wedge^2 V$ to denote the exterior square of $V$; in other words, $\wedge^2 V$ is the two-index antisymmetric tensor representation.} that $\so(V) \cong \wedge^2 V$ as $\so(V)$-representations and that $\wedge^2 V_7 \cong V_7 \oplus \fg_2$ as $\fg_2$-representations, it is straightforward to decompose the dg Lie algebra $(\fp, [Q, -])$. Explicitly, it takes the following form:
\[
	\begin{tikzcd}[row sep = tiny]
	V_7 \arrow[r] &	V_7 \otimes \textbf{1}^{-1} & \\
	\fg_2&	 V_7 \otimes \textbf{1}^1 \arrow[r] &		V_7\\
	V_7 \otimes \textbf{2}^1 \arrow[r] &	V_7 \otimes \textbf{2}^0 & \textbf{2}^1\\
	V_7 \otimes \textbf{2}^{-1} & \textbf{2}^0 \arrow[r] & \textbf{2}^{-1} \\
	\textbf{1}^0 \arrow[r] & \textbf{1}^{-1}\\
	\textbf{1}^{2} \arrow[r] & \textbf{1}^1 \\
	\textbf{1}^{-2} \\
	\mathfrak{sl}(2). \\
	\end{tikzcd}
\]
Note that the arrows denote the action of the differential $[Q,-]$. Since each arrow constitutes an equivariant map between finite-dimensional irreducible representations of $G_2 \times \SU(2)$, Schur's lemma implies that these maps are simply multiples of the identity. It is thus immediate to read off the cohomology
\[
	H^\bu(\fp,[Q,-]) = \big( \fg_2 \oplus \mathfrak{sl}(2) \oplus V_7 \otimes \textbf{2}^{-1} \oplus \textbf{1}^{-2} \big) \ltimes L .
\]

\subsubsection[Homotopy transfer and L\_infty structure]{Homotopy transfer and $\boldsymbol{ L_\infty}$ structure}\label{p: maxtransfer}
Let us now fix a retraction
\begin{equation}
\begin{tikzcd}
\arrow[loop left]{l}{h}(\fp , [Q,-])\arrow[r, shift left, "p"] &(H^\bu(\fp,[Q,-]) , 0)\arrow[l, shift left, "i"] \:
\end{tikzcd}
\label{eq: hotop data}
\end{equation}
and perform homotopy transfer of the Lie structure on $(\fp , [Q,-])$ along this diagram. This equips the residual super-Poincar\'e algebra $\fp_Q$ with the structure of an $L_\infty$ algebra.

Explicitly, we can fix the homotopy data~\eqref{eq: hotop data} by setting $i$ and $p$ to be the obvious inclusion of and projection to the cohomology. Further, we can take $h$ to be the inverse of $[Q,-]$ along all the arrows in~\eqref{eq: hotop data} and vanishing everywhere else.

The Lie bracket on $\fp$ induces a binary bracket on $\fp_Q$ in straightforward manner. Explicitly, this equips $\fg_2$ and $\lie{sl}(2)$ with their standard Lie brackets and lets them act on the other summands according to their representations. In addition, the bracket between two elements $V_7 \otimes L^\vee$ maps to $\textbf{1}^{-2}$ by using the trace on the two elements in $V_7$ and the isomorphism $\wedge^2 L^\vee \cong \textbf{1}^{-2}$.

Higher brackets arising from homotopy transfer are obtained by sum-over-trees formulas~\cite{LodayVallette}. In our case, an additional four-ary bracket arises from diagrams of the following form:
\begin{equation} \label{eq: mu4-diags}
\begin{split}& \text{$\begin{tikzpicture}
\begin{feynman}
\vertex at (-2,-0.5) {$\mu_4 \ = $};
\vertex(a) at (-1,1) {$i$};
\vertex(b) at (-1,0) {$i$};
\vertex(c) at (-1,-1) {$i$};
\vertex(c2) at (-1,-2) {$i$};
\vertex(d) at (0,0.5);
\vertex(e) at (1,0);
\vertex(e2) at (2,-0.5);
\vertex(f) at (3,-0.5) {$p$};
\diagram* {(a)--(d), (b)--(d), (d)--[edge label = $h$](e), (c)--(e), (c2)--(e2), (e)--[edge label = $h$](e2), (e2)--(f)};
\vertex(a') at (5,1) {$i$};
\vertex(b') at (5,0) {$i$};
\vertex(c') at (5,-1) {$i$};
\vertex(c2') at (5,-2) {$i$};
\vertex(d') at (6,0.5);
\vertex(e') at (7,0);
\vertex(e2') at (8,-0.5);
\vertex(f') at (9,-0.5) {$p$};
\diagram*{(a')--(d'), (c')--(d'), (d')--[edge label = $h$](e'), (b')--(e'), (c2')--(e2'), (e')--[edge label = $h$](e2'), (e2')--(f')};
\vertex at (4,-0.5) {$+$};
\vertex at (10.25,-0.5) {$+ \cdots.$};
\end{feynman}
\end{tikzpicture}$}\end{split}
\end{equation}
Here, the binary vertices mean application of the binary bracket in $\fp$ while the homotopy is applied at each internal line. In formulas, this means that we have
\[
 \mu_4(x_1, x_2, x_3 , x_4) = p[i(x_4) , h[i(x_3) , h[i(x_2) , i(x_1)] ] ] + \cdots.
\]
One has to evaluate all such diagrams using our homotopy data in order to see what contributions can arise. This can be done systematically by decomposing the Lie bracket in $\fp$ under $G_2 \times \SU(2)$; more details can be found in Appendix~\ref{ap: details}.
In our case, one finds that there are two non-vanishing contributions to $\mu_4$. Both take three-inputs in degree zero and one input in degree two, i.e.,
\[
 \mu_4 \colon\ (\fp_Q)_2 \otimes [(\fp_Q)_0]^{\otimes 3} \longrightarrow (\fp_Q)_0.
\]
The first, is given by a map
\[
 \mu_4^{(A)} \colon\ \textbf{2}^{1} \otimes \big(V_7 \otimes \textbf{2}^{-1}\big)^{\otimes 3} \longrightarrow \fg_2 \oplus \lie{sl}(2),
\]
while the second option is described by
\[
 \mu_4^{(B)}\colon\ \textbf{2}^1 \otimes \textbf{1}^{-2} \otimes \big(V_7 \otimes \textbf{2}^{-1}\big)^{\otimes 2} \longrightarrow V_7 \otimes \textbf{2}^{-1}.
\]

We start by investigating $\mu_4^{(A)}$.
Taking into account the antisymmetry of the four-ary bracket, it is easy to see on purely representation theoretic grounds that the map is unique up to scalar multiples. Indeed, decomposing $\textbf{2}^1 \otimes \wedge^3\big(V_7 \otimes \textbf{2}^{-1}\big)$ into irreducibles, we find a single copy of~${\mathfrak{sl}(2) \oplus \fg_2}$. We denote our inputs by $v \in L \cong \textbf{2}^1$ and $R_1,R_2,R_3 \in V_7 \otimes \textbf{2}^{-1} = V_7 \otimes L^\vee$ and expand $R_i = r_i \otimes f_i^\vee$ we find
\begin{align*}
 \mu_4^{(A)}(v, R_1, R_2 , R_3)
 ={}& \varphi(r_1,r_2,r_3) \big(f_1^\vee \odot f_2^\vee \odot f_3^\vee\big)(v) \\
 &+ \sum_{\sigma \in S_3} p_{\fg_2}( r_{\sigma(3)} \wedge p_{V_7}(r_{\sigma(1)} \wedge r_{\sigma(2)})) \big(f^\vee_{\sigma(2)} \wedge f^\vee_{\sigma(3)}\big) f^\vee_{\sigma(1)}(v).
\end{align*}
The notation in this formula deserves some explanation. First, $\varphi$ denotes the $G_2$ three-form on $V_7$ while $\odot$ are symmetrized tensor products. This means that the first term can be viewed as an element in $\mathfrak{sl}(2)$ by identifying the adjoint representation with the symmetric square of the fundamental representation ($\mathfrak{sl}(2) \cong \textbf{3})$. To understand the second term, recall the decomposition
$
 \wedge^2 V_7 \cong V_7 \oplus \fg_2 $.
We denote the projections on these factors by $p_{V_7}$ and $p_{\fg_2}$. Note that these projectors can be expressed in terms of $\varphi$ as
$
 p_{V_7}(r) = (\varphi \vee r)^\sharp$ and $ p_{\fg_2} = \id - p_{V_7}$.
Here and in the following we denote the isomorphisms $V^\vee \cong V$ induced by the metric by raising and lowering indices with superscripts $^\sharp$ and $^\flat$. Further, $\vee$ denotes contraction between a form and a vector.

Let us now turn our attention to \smash{$\mu_4^{(B)}$}.
Again, we see on representation theoretic grounds that there is a unique such map up to a prefactor. We find
\[
 \mu_4^{(B)}(v, \alpha, R_1, R_2) = \alpha \cdot \big(f_1^\vee \wedge f_2^\vee\big) (r_1 \vee r_2 \vee \varphi)^\sharp \otimes v ,
\]
where $\alpha$ is the input in $\textbf{1}^{-2}$.

The full four-ary bracket is then the sum of both terms, \smash{$\mu_4 = \mu_4^{(A)} + \mu_4^{(B)}$}. Writing down possible trees for operations with more than four external legs, it is easy to see that there can't be any additional higher brackets $\mu_k$ for $k>4$. Thus, $(\fp_Q , \mu_2 = [-,-] , \mu_4)$ completely specifies the residual super-Poincar\'e algebra as an $L_\infty$ algebra.

\subsection{A module structure for the minimal model} \label{sec: max-module}
The residual super-Poincar\'e $L_\infty$ algebra $\fp_Q$ acts on the twisted theory. In the following we will describe this module structure, first on the minimal model, then on the component fields.

\subsubsection{The minimal model of maximally twisted eleven-dimensional supergravity}
As described in~Section~\ref{p: first-max}, the maximal twist in a flat background is Poisson--Chern--Simons theory on $\RR^7 \times \CC^2$. The minimal model of this $L_\infty$ algebra is holomorphic functions on $\CC^2$,
$ \big( \cO\big(\CC^2\big), \{-,-\} \big)$.
There is only the binary operation, again given by the Poisson bracket. In coordinates, it takes the usual form
\[
 \{f_1, f_2\} = \frac{\partial f_1}{\partial z_1} \frac{\partial f_2}{\partial z_2} - \frac{\partial f_2}{\partial z_1} \frac{\partial f_1}{\partial z_2} .
\]
We further recall that $\bigl(\cO\big(\CC^2\big), \{- , -\}\bigr)$ is a central extension of divergence-free vector fields,
\[
 0 \longrightarrow \CC \longrightarrow \cO\big(\CC^2\big) \longrightarrow \Vect_0\big(\CC^2\big) \longrightarrow 0 .
\]
This is done by assigning to a holomorphic function $f \in \cO\big(\CC^2\big)$ its Hamiltonian vector field $X_f = \{f,-\}$.

\subsubsection{The action on the minimal model}
The $L_\infty$ algebra $\fp_Q$ acts on the minimal model $\big(\cO\big(\CC^2\big) , \{-,-\}\big)$ turning it into a $\fp_Q$-module. Conceptually, one should think of this module structure as arising from the action of the full super-Poincar\'e algebra on the BV fields of the eleven-dimensional supergravity multiplet via homotopy transfer of module structures. Here, instead of deriving the module structure via homotopy transfer, we take a more direct route and solve the $L_\infty$ relations directly.

An inner action is given by an $L_\infty$ map\footnote{Here and for the following discussions of module structures, we view the right hand side as an $L_\infty$ algebra by taking global sections of the underlying local $L_\infty$ algebra. Alternatively, we can also view the left hand side as a~local $L_\infty$ by promoting it to the locally constant sheaf on $\CC^2$ and turning $\rho$ into a map of local $L_\infty$ algebras.}
$
 \rho\colon (\fp_Q , \mu_2, \mu_4) \rightsquigarrow \big(\cO\big(\CC^2\big) , \{-,-\}\big)$.
We can view this as an action of vector fields by identifying holomorphic functions with its Hamiltonian vector field.

The strict part of the module structure is fixed by the geometric action of holomorphic translations and $\lie{sl}(2)$ on $\CC^2$. Explicitly, we want these to act by the usual vector fields
$
 v^i \partial_{z_i}$, $v \in L $ and $ A_{ij}z^i \partial_{z_j}$, $ A \in \mathfrak{sl}(2)$.
Fixing a basis $e_i$ for $L$ and $(e,h,f)$ for $\lie{sl}(2)$, this is achieved by setting
\begin{gather}
 \rho^{(1)}(e_1) = -z_2, \qquad \rho^{(1)}(e_2) = z_1, \nonumber\\
 \rho^{(1)}(e) = \frac{1}{2}(z_1)^2, \qquad \rho^{(1)}(h) = -z_1 z_2, \qquad \rho^{(1)}(f) = \frac{1}{2}(z_2)^2. \label{eq: rho1}
\end{gather}
All other elements act trivially at the strict level, i.e., the kernel of $\rho^{(1)}$ is $\fg_2 \oplus \big(V_7 \otimes \textbf{2}^{-1}\big) \oplus \textbf{1}^{-2}$.

In order to identify any higher operations for our module structure, we solve the $L_\infty$ coherence relation. The first nontrivial relation arises with four inputs and simplifies to
\[
 \big\{ \rho^{(1)}(v) , \rho^{(3)} (R_1,R_2,R_3) \big\} = \rho^{(1)}(\mu_4(v,R_1,R_2,R_3)).
\]
We immediately see that the term on the right-hand-side is given by the quadratic polynomials representing $\mathfrak{sl}(2)$ so that we can solve this relation for $\rho^{(3)}$ to find
\[
 \rho^{(3)}(R_1, R_2, R_3) = \frac{1}{6} \varphi(r_1,r_2,r_3) \big(f_1^\vee \odot f_2^\vee \odot f_3^\vee\big)^{ijk} z_i z_j z_k.
\]
Here and in the following equations we continue to use the notation $R_i=r_i \otimes f_i^\vee$ for inputs in~${V_7 \otimes L^\vee}$.

This means the image of $\rho^{(3)}$ consists of all cubic polynomials in $\cO\big(\CC^2\big)$.
It is easy to compute the corresponding Hamiltonian vector fields
\begin{gather}
 X_{\rho^{(3)}(r_1 \otimes e_1^\vee,r_2 \otimes e_1^\vee,r_3 \otimes e_1^\vee)} = \frac{1}{2} \varphi(r_1,r_2,r_3) (z_1)^2 \frac{\partial}{\partial z_2},\nonumber \\
 X_{\rho^{(3)}(r_1 \otimes e_1^\vee, r_2 \otimes e_1^\vee, r_3 \otimes e_2^\vee)} = \frac{1}{2} \varphi(r_1,r_2,r_3) \left[ (z_1)^2 \frac{\partial}{\partial z_1} - 2z_1 z_2 \frac{\partial}{\partial z_2} \right],\nonumber\\
 X_{\rho^{(3)}(r_1 \otimes e_2^\vee, r_2 \otimes e_2^\vee, r_3 \otimes e_1^\vee)} = \frac{1}{2} \varphi(r_1,r_2,r_3) \left[ (z_2)^2 \frac{\partial}{\partial z_2} -2z_1 z_2 \frac{\partial}{\partial z_1} \right],\nonumber\\
 X_{\rho^{(3)}(r_1 \otimes e_2^\vee,r_2 \otimes e_2^\vee,r_3 \otimes e_2^\vee)} = \frac{1}{2} \varphi(r_1,r_2,r_3) (z_2)^2 \frac{\partial}{\partial z_1}. \label{eq: rho3}
\end{gather}

Moving on, the bracket between two $\rho^{(3)}$'s gives rise to the $L_\infty$ relation with six inputs. This relation can be simplified to
\begin{gather*}
 \frac{1}{2} \sum_{\sigma \in \mathrm{Sh}(3;6)} \chi(\sigma) \big\{ \rho^{(3)}(R_{\sigma(1)},R_{\sigma(2)},R_{\sigma(3)}) , \rho^{(3)} (R_{\sigma(4)},R_{\sigma(5)},R_{\sigma(6)}) \big\} \\
\qquad = \sum_{\sigma \in \mathrm{Sh}(2;6)} \chi(\sigma) \rho^{(5)}( \mu_2(R_{\sigma(1)},R_{\sigma(2)}),R_{\sigma(3)},R_{\sigma(4)},R_{\sigma(5)} , R_{\sigma(6} ),
\end{gather*}
where $\mathrm{Sh}(i;n)$ denotes the set of $(i,n)$-shuffles\footnote{Recall that an $(i,n)$-shuffle is a permutation $\sigma \in S_n$ such that the first $i$ and the last $n-i$ elements are ordered, i.e., $\sigma(1) < \sigma(2) < \dots< \sigma(i)$ and $\sigma(i+1) < \sigma(i+2) < \dots < \sigma(n)$.} and $\chi(\sigma)$ is the Koszul-sign of the permutation~$\sigma$.
Remembering that the binary bracket on $\fp_Q$ maps $\wedge^2 \big(V_7 \otimes L^\vee\big) \rightarrow \textbf{1}^{-2}$, this means that there is a five-ary term in the module structure of the form
\[
 \rho^{(5)} \colon\ \textbf{1}^{-2} \otimes \wedge^4 \big(V_7 \otimes L^\vee\big) \longrightarrow \cO\big(\CC^2\big).
\]
The image of $\rho^{(5)}$ are quartic polynomials, explicitly we find
\[
 \rho^{(5)}(\alpha, R_1, R_2, R_3 , R_4) = \alpha (\star \varphi)(r_1,r_2,r_3,r_4) \big(f_1^\vee \odot f_2^\vee \odot f_3^\vee \odot f_4^\vee\big)^{ijkl} z_i z_j z_k z_l,
\]
where $\star \varphi$ denotes the Hodge dual of the $G_2$ three form $\varphi$. Again, it is easy to compute the associated Hamiltonian vector fields. Explicitly, the image of $\rho^{(5)}$ is spanned by
\[
 (z_1)^3 \frac{\partial}{\partial z_2}, \qquad (z_1)^2 z_2 \frac{\partial}{\partial z_2} - \frac{1}{3} (z_1)^3 \frac{\partial}{\partial z_2}, \qquad z_1 (z_2)^2 \frac{\partial}{\partial z_2} - (z_2)^2 z_1 \frac{\partial}{\partial z_1}, \qquad z_1 \leftrightarrow z_2.
\]
This fixes the $L_\infty$ map completely. One can check on purely representation theoretic grounds that the brackets that could contribute to higher term relations all vanish. For example, terms like $\big\{\rho^{(3)} , \rho^{(5)} \big\}$ appear in the $L_\infty$ relation with eight inputs. The result of this bracket would be a polynomial of fifth order living in the representation $\Sym^5\big(\CC^2\big) \cong \textbf{6}$. However, this relation is completely antisymmetric on the inputs, so that this term would represent a map $\textbf{1} \otimes \wedge^7 \big({V_7 \otimes L^\vee}\big) \rightarrow \textbf{6}$. It is easy to check from the decomposition of the source into irreducible representations that there are no such $G_2 \times \SU(2)$ maps except zero. A similar argument applies to brackets of the form $\big\{\rho^{(5)}, \rho^{(5)} \big\}$ appearing in the $L_\infty$ relation with ten inputs.

\subsubsection{The action on component fields}
The above action of $\fp_Q$ on $\cO\big(\CC^2\big)$ can be extended to an action on the component fields $\Omega^{0,\bu}\big(\CC^2\big) \otimes \Omega^\bu\big(\RR^7\big)$ in an obvious way. This extension is given by a $L_\infty$ map
\[
 \Tilde{\rho}\colon\ \fp_Q \rightsquigarrow \big(\Omega^{(0,\bu)}\big(\CC^2\big) \otimes \Omega^\bullet\big(\RR^7\big) , \bar{\partial} + \d , \{-,-\} \big).
\]
Note that $\fp_Q$ has vanishing differential so that $\Tilde{\rho}{(1)}$ has to satisfy
\smash{$
 \big(\d + \bar{\partial}\big) \circ \Tilde{\rho}^{(1)} = 0$}.
This means that elements in the image of \smash{$\Tilde{\rho}^{(1)}$} have to be constant along the topological directions and depend holomorphically on $\CC^2$. Including holomorphic functions into zero forms along the obvious map~${i \colon \cO\big(\CC^2\big) \hookrightarrow \Omega^{(0,0)}\big(\CC^2\big) \otimes \Omega^0\big(\RR^7\big)}$, we find that $\Tilde{\rho} = i \circ \rho$ defines an $L_\infty$ module structure on the component fields.

\subsubsection{The action in terms of the physical fields}
In Section~\ref{p: max-stair}, we sketched how the fields in the maximal twist correspond to field configurations in the physical theory. In the physical theory, we know that the even piece of the supersymmetry algebra acts by vector fields through the diffeomorphism ghost.

Indeed, we can compose $\tilde{\rho}$ with the embedding of the $\Omega^{(0,0)}\big(\CC^2\big) \otimes \Omega^0\big(\RR^7\big)$ to the fields of the physical configurations along the representatives determined by the stair in~\eqref{eq: stair}. Note that a~holomorphic function $f(z)$ thereby gives rise to a value of the diffeomorphism ghost $
 \xi = \Omega^{-1} \vee \partial f $.
In other words, the diffeomorphism ghost is identified with the Hamiltonian vector field of~$f$. Consequently, we recover the vector fields~\eqref{eq: rho1} and~\eqref{eq: rho3} as field configurations of the diffeomorphism ghost in the image of the composition
\[
 \fg_Q \rightsquigarrow (\Omega^\bu\big(\CC^2\big) \otimes \Omega^\bu\big(\RR^7\big) , \d + \bar{\partial} , \{-,-\} ) \hookrightarrow (\cE_{\rm BV} , Q_{\rm BV} + Q) .
\]

\subsubsection{Summary}
The four-ary brackets constructed in this section show that the residual super-Poincar\'e algebra in maximally twisted supergravity should be viewed as an $L_\infty$ algebra rather than a super Lie algebra. Further, these symmetries are represented in non-trivial ways both on the minimal model and the component fields of the twisted theory. Interestingly, even at the level of the minimal model in a flat background where all the topological directions have been integrated out, the action of the residual symmetries depends on the $G_2$ three-form, suggesting that through these higher-order terms, the twisted four-dimensional holomorphic field theory remembers something of its M-theoretic origins. It would be interesting to see how $\fp_Q$, the minimal model for the BV-BRST fields of the maximal twist, and the action of the former on the latter, are modified when~$\R^7$ is replaced with a nontrivial $G_2$-manifold. In that case Kaluza--Klein compactification gives a~holomorphically twisted four-dimensional $\mathcal{N}=1$ theory on the remaining $\C^2$, which will admit currents realizing the action of residual symmetries in supergravity. Further, reducing on a~circle in the holomorphic directions gives an A-type topological twist of seven-dimensional~${\mathcal{N}=1}$ whose equations of motion count $G_2$-instantons \cite{EvidenceG2}. It would be interesting to see if our residual symmetries imply any higher structure on the conjectured algebra of $G_2$-instantons, thought to model BPS states in this seven-dimensional theory.

Moreover, in \cite{CostelloMtheory2}, Costello defines a class of $\Omega$-backgrounds for the maximal twist, in settings where the $G_2$-manifold has an isometry with isolated fixed points; this background should lead to further deformations of the residual symmetry algebra.

\section{Residual symmetries in the minimal twist} \label{sec: min}
We now turn our attention to the minimal twist of eleven-dimensional supergravity. In~Section~\ref{sec: min-transfer}, we construct the higher $L_\infty$ operations on the residual super-Poincar\'e algebra, and in~Section~\ref{sec: min-module}, we describe a module structure on the minimally twisted supergravity theory. Finally, in~Section~\ref{sec: further}, we show how the maximal twist arises as a further twist from the minimal twist.

\subsection{The residual super-Poincar\'e algebra and homotopy transfer} \label{sec: min-transfer}
Let us now fix a minimal twisting supercharge $Q$. The choice of $Q$ fixes a maximally isotropic subspace $L\subset V$ and thereby a decomposition
$
 V \cong L \oplus L^\vee \oplus \CC$,
into holomorphic, anti-holomorphic and one topological direction. In order to study the residual super-Poincar\'e algebra, we decompose $(\fp, [Q,-])$ into $\lie{sl}(5)$ representations. This decomposition and the description of $\fp_Q$ as a~super Lie algebra was already done in~\cite{spinortwist}
\[
 \begin{tikzcd}[row sep = tiny]
 L^\vee & \wedge^5 L \arrow[r] & \CC \\
 \wedge^2 L^\vee & \wedge^4 L \arrow[r] & L^\vee \\
 \mathfrak{sl}(5) & \wedge^3 L & L \\
 \wedge^2 L \arrow[r] & \wedge^2 L \\
 L \arrow[r] &\wedge^1 L \\
 \mathfrak{gl}(1) \arrow[r] & \wedge^0 L.
	\end{tikzcd}
\]
We can read off the cohomology to find
\[
 \fp_Q = \mathfrak{sl}(5) \ltimes \big(L^\vee \oplus \wedge^2 L^\vee\big) \oplus \wedge^3 L(-1) \oplus L (-2).
\]
Again, the Lie bracket on the super-Poincar\'e algebra descends to a binary bracket on $\fp_Q$. This equips $\mathfrak{sl}(5)$ with its standard Lie bracket; further $\mathfrak{sl}(5)$ acts on all other summands according to their representation. In addition, there is a bracket $L^\vee \times L^\vee \rightarrow \wedge^2 L^\vee$ in the obvious way.

Note that, as opposed to the maximal twist, there are odd elements in the residual super-Poincar\'e algebra. It is convenient to apply the identifications $\wedge^3 L \cong \wedge^2 L^\vee$. Then, the bracket is given by the wedge product followed by the identification $L \cong \wedge^4 L^\vee$.
This also means that~$\fp_Q$ has a non-trivial nilpotence variety encoding further twists; indeed, the $Y_Q$ has a single orbit corresponding to the maximal twist viewed as a further twist of the minimal twist and is isomorphic to the affine cone over the Grassmanian $\mathrm{Gr}(2,5)$~\cite{spinortwist}.

Homotopy transfer gives rise to three-ary and four-ary brackets. These terms can be systematically identified by writing all possible diagrams contributing to the homotopy transfer procedure using the equivariant decomposition of the bracket in $\fp$.

There are two types of terms contributing to $\mu_3$. The first takes two inputs from degree zero and one input from degree two so that the output is in degree one. Explicitly, the homotopy transfer procedure induces a map
\[
 \mu_3^{(A)}\colon\ L(-2) \otimes \wedge^2 L^\vee \otimes L^\vee \longrightarrow \wedge^3 L (-1) .
\]
There are two diagrams that contribute to $\mu_3^{(A)}$
\[
\begin{tikzpicture}
\begin{feynman}
\vertex at (-3,0) {$\mu_3^{(A)} \ = $};
\vertex(a) at (-1.5,1) {$L$};
\vertex(b) at (-1.5,0) {$\wedge^2 L^\vee$};
\vertex(c) at (-1.5,-1) {$L^\vee$};
\vertex(d) at (0,0.5);
\vertex(e) at (1,0);
\vertex(f) at (2,0) {$p$};
\vertex at (2.8,0) {$+$};
\vertex(a') at (4,1) {$L$};
\vertex(b') at (4,0) {$\wedge^2 L^\vee$};
\vertex(c') at (4,-1) {$L^\vee$};
\vertex(d') at (5.5,0.5);
\vertex(e') at (7,0);
\vertex(f') at (8.5,0) {$p$,};
\diagram* {(a)--(d), (b)--(d), (d)--[edge label = $h$](e), (c)--(e), (f)--(e)};
\diagram* {(a')--(d'), (c')--(d'), (d')--[edge label = $h$](e'), (b')--(e'), (f')--(e')};
\vertex at (8.5,0);
\end{feynman}
\end{tikzpicture}
\]
where we already denoted the inputs such that a non-vanishing contribution arises.
Explicitly for inputs $v \in L$, $B \in \wedge^2 L^\vee$, and $R \in L^\vee$, we have
$
 \mu_3(v,B,R) = R \wedge (v \vee B) + B \cdot R(v)$,
where we leave the isomorphism $\wedge^2 L^\vee \cong \wedge^3 L$ implicit.

The second takes two inputs in degree zero and one input degree one so that the output lies in degree zero,
\[
 \mu_3^{(B)}\colon\ \wedge^3 L(-1) \otimes \wedge^2 L^\vee \otimes L^\vee \longrightarrow \mathfrak{sl}(5).
\]
Again, there are two diagrams contributing to $\mu_3^{(B)}$
\[
\begin{tikzpicture}
\begin{feynman}
\vertex at (-3,0) {$\mu_3^{(B)} \ = $};
\vertex(a) at (-1.5,1) {$\wedge^3 L$};
\vertex(b) at (-1.5,0) {$\wedge^2 L^\vee$};
\vertex(c) at (-1.5,-1) {$L^\vee$};
\vertex(d) at (0,0.5);
\vertex(e) at (1,0);
\vertex(f) at (2,0) {$p$};
\vertex at (2.8,0) {$+$};
\vertex(a') at (4,1) {$\wedge^3 L$};
\vertex(b') at (4,0) {$L^\vee$};
\vertex(c') at (4,-1) {$\wedge^2 L^\vee$};
\vertex(d') at (5.5,0.5);
\vertex(e') at (7,0);
\vertex(f') at (8.5,0) {$p$.};
\diagram* {(a)--(d), (b)--(d), (d)--[edge label = $h$](e), (c)--(e), (f)--(e)};
\diagram* {(a')--(d'), (c')--(d'), (d')--[edge label = $h$](e'), (b')--(e'), (f')--(e')};
\vertex at (8.5,0);
\end{feynman}
\end{tikzpicture}
\]
For inputs $q \in \wedge^3 L$, $B \in \wedge^2 L$, and $R \in L^\vee$, we find the following form:
\[
 \mu_3(q,B,R) = p_{\mathfrak{sl}(5)} ( R \otimes (q \vee B)) + p'_{\mathfrak{sl}(5)} (B \otimes (q \vee R)),
\]
where \smash{$p_{\mathfrak{sl}(5)}$} and \smash{$p'_{\mathfrak{sl}(5)}$} are the projections $L \otimes L^\vee \rightarrow \mathfrak{sl}(5)$ (given by removing the trace) and~${\wedge^2 L \otimes \wedge^2 L^\vee \rightarrow \mathfrak{sl}(5)}$, respectively.

Finally, the four ary bracket is given by a map
\[
 \mu_4 \colon\ L(-2) \otimes \wedge^3\big(\wedge^2 L^\vee\big) \longrightarrow \mathfrak{sl}(5)
\]
that arises from the diagrams~\eqref{eq: mu4-diags}. Explicitly, the four-ary bracket yields
\[
 \mu_4(v,B_1,B_2,B_3) = \sum_{\sigma \in S_3} \chi(\sigma) p'_{\mathfrak{sl}(5)} \left[ B_{\sigma(3)} \otimes \big( B_{\sigma(2)} \vee \Omega^{-1} (v \vee B_{\sigma(1)}) \big) \right],
\]
where the inverse of the holomorphic volume form $\Omega^{-1}$ provides an isomorphism $L^\vee \cong \wedge^4 L$.

Again, explicitly writing out diagrams with more than four external legs, one finds that there are no higher brackets $\mu_k$ for $k>4$.

\subsection{A module structure for the minimal model} \label{sec: min-module}
Let us now move on to discuss the module structure of this $L_\infty$ algebra on the fields of minimally twisted supergravity.

\subsubsection{The minimal model of minimally twisted eleven-dimensional supergravity}
In~\cite{RSW}, the minimal model of the minimally twisted theory was identified with a central extension of the infinite-dimensional exceptional super Lie algebra $E(5|10)$. This description is obtained by computing the cohomology of the component fields~\eqref{eq: comp-min-tw} and performing the homotopy transfer of the $L_\infty$ structure. The even piece of $E(5|10)$ consists of divergence-free vector fields, while the odd piece is identified as closed two-forms:
$
 E(5|10)_+ = \Vect_0\big(\CC^5\big) $, $ E(5|10)_- = \Omega^2_{cl}\big(\CC^5\big) $.
The bracket between two even elements is just the bracket of vector fields, the bracket between a~vector field $X$ and a two-form $\alpha$ is given by the Lie derivative
$
 [X,\alpha] = \cL_X \alpha$,
finally, the bracket between two odd elements is defined by
$
 [\alpha, \beta] = \iota_{\Omega^{-1}} (\alpha \wedge \beta)$.
Here, $\Omega = \d z_1 \wedge \dots \wedge \d z_5$ denotes the holomorphic volume form so that $\Omega^{-1}$ is the corresponding polyvector $\partial_{z_1} \wedge \dots \wedge \partial_{z_5}$.

The central extension introduces an additional copy of $\C$ whose elements we denote by $b$. The corresponding higher bracket is
\[ 
 \mu_3\colon\ E(5|10)^{\otimes 3} \longrightarrow \C_b, \qquad \mu_3(\alpha, X, X') = \alpha(X \wedge X')(0) ,
\]
i.e., we evaluate the two-form $\alpha$ on a pair of vectors at the origin $0 \in \CC^5$. We denote the centrally extended algebra by \smash{$\widehat{E(5|10)}$}.

\subsubsection{The action on the minimal model}
The action on the minimal model is given by an $L_\infty$ map $\fp_Q \rightsquigarrow \smash{\widehat{E(5|10)}}$. In \cite{RSW}, the authors considered the problem of constructing an inner action of the strict Lie algebra underlying~$\mathfrak{p}_Q$ on~\smash{$\widehat{E(5|10)}$}. They found that one needs to consider a certain one-dimensional $L_\infty$ central extension of this strict Lie algebra. The $L_\infty$ extension considered therein, is of a different nature than those we consider~-- their presence is owed to the fact that eleven-dimensional supergravity is more properly thought of as a gauge theory for an $L_\infty$ algebra denoted $\mathfrak{m2brane}$ that can be thought of as extending the super-Poincar\'e algebra by the tower of gauge transformations for the $C$-field. In addition to the usual Lie brackets on the super-Poincar\'e algebra, $\mathfrak{m2brane}$ features a single additional four-ary operation that takes two odd elements and two translations to the one-dimensional central summand. Such central extensions of super-Poincar\'e algebras in ten and eleven dimensions have a rich history (see, for example, \cite{FSS}); they correspond to BPS branes and are often dubbed ``brane scan'' cocycles. The resulting $L_\infty$ algebras provide local models of the Lie $n$-algebroids underlying generalized tangent bundles featuring in descriptions of non-geometric supergravity backgrounds via generalized geometry \cite{Cederwall_2019, malgebra,Hull_2007, smithwaldram}.

The authors of \cite{RSW} computed the $Q$-twist of $\mathfrak{m2brane}$ keeping track of only those higher brackets generated by the four-ary bracket, and found the result to be an $L_\infty$ algebra with a~three-ary bracket. The resulting $L_\infty$-algebra has a strict map to \smash{$\widehat{E(5|10)}$}. In this paper, we will produce an $L_\infty$ map
$\rho\colon\mathfrak{p}_Q \rightsquigarrow E(5|10)$
leaving questions of how the $L_\infty$ brackets on~$\mathfrak{p}_Q$ interact with $L_\infty$ brackets obtained by twisting brane scan cocycles, along with potential applications to descriptions of non-geometric supergravity backgrounds, to future work.

The strict part of the map is easily described: Rotations $A \in \mathfrak{sl}(5)$ and holomorphic translations $v \in L$ are mapped to the corresponding vector fields
\smash{$
 \rho^{(1)}(v) = v^i \partial_{z^i} $}, \smash{$
 \rho^{(1)}(A) = A_{i}^{\ j} z^i \partial_{z^j} $},
while we set for the odd elements $q \in \wedge^2 L^\vee$
\smash{$
 \rho^{(1)}(q) = q_{ij} \d z^i \wedge \d z^j$}.
The $L_\infty$ relation with two inputs has two contributions that have to satisfied, namely one coming from \smash{$\mu_3^{(A)}$} and one from~\smash{$\mu_3^{(B)}$}. These read
\[
 \big[\rho^{(1)}(v) , \rho^{(2)} (R,B)\big]_{E(5|10)} = \rho^{(1)}(\mu_3(v,R,B))
\]
and
\[
 \big[\rho^{(1)}(q) , \rho^{(2)}(R,B)\big]_{E(5|10)} = \rho^{(1)}(\mu_3(q,R,B)).
\]
In the first relation, recall that the bracket on the left-hand-side of this relation is given by the Lie derivative, while in the second relation it arises from the wedge product and the contraction with $\Omega^{-1}$. In both cases, we see that $\rho^{(2)}(R,B)$ is a two-form with a coefficient linear in $z_i$. Solving these relations explicitly gives
$
 \rho^{(2)}(B,R) = (B_{ij} R_k + R_{[i} B_{j]k} ) z^k \d z^i \wedge \d z^j$.

Next, we move on to the $L_\infty$ relation with four inputs. Again, there are two different contributions. First, the four-ary bracket $\mu_4$ gives rise to
\[
 \big[\rho^{(1)}(v), \rho^{(3)}(B_1,B_2,B_3)\big]_{E(5|10)} = \rho^{(1)}(\mu_4(v,B_1,B_2,B_3)).
\]
In addition, the bracket between two $\rho^{(2)}$'s also contributes. In our case, this part of the relation can be simplified to
\[
 \big[\rho^{(2)}(R_1, B_1) , \rho^{(2)}(R_2,B_2)\big]_{E(5|10)} = \rho^{(3)}(\mu_2(R_1,R_2), B_1,B_2).
\]
Recall that the $\mu_2$ on the right-hand-side of this equation maps the two inputs $R_1, R_2 \in L^\vee$ to~${R_1 \wedge R_2 \in \wedge^2 L^\vee}$.

From both equations, we find that elements in the image of $\rho^{(3)}$ are divergence-free vector fields on $\CC^5$ with coefficients that are quadratic in the coordinates $z_i$. This means that $\rho^{(3)}$ is a map
\smash{$
 \rho^{(3)}\colon \wedge^3\big(\wedge^2 L^\vee\big) \longrightarrow \Vect_0\big(\CC^5\big)$}.
It is easy to check that there is a unique such $\SU(5)$-equivariant map that lands in quadratic vector fields and solves both relations. Explicitly, we can write
\[
 \rho^{(3)}(B_1,B_2,B_3) = p(B_1 \wedge B_2 \wedge B_3)^{ij}_k z_i z_j \frac{\partial}{\partial z_k},
\]
where $p$ is the projection $\wedge^3\bigl(\wedge^2 L^\vee\bigr) \rightarrow [1,0,0,2]$.
Again, one can investigate the terms contributing to the higher term relations on purely representation theoretic grounds to find that there are no additional contributions to the higher $L_\infty$ relations with more inputs, showing that this discussion fixes the $L_\infty$ module structure completely.

\subsubsection{The action on component fields}
As for the maximal twist, we expect to obtain an action of $\fp_Q$ on component fields of the field strength formulation of the minimal twist from~\eqref{eq: comp-min-tw-str} by postcomposing the map $\rho\colon \fp_Q\rightsquigarrow E(5|10)$ with a suitable inclusion from $E(5|10)$ to the component fields. Such an inclusion is part of a~homotopy retract between the component fields and the minimal model and is given by the obvious inclusion of holomorphic divergence free vector fields and holomorphic closed two forms.

\subsection{The maximal twist as a further twist} \label{sec: further}
We can pick an odd square-zero element $Q_{\max}$ in the residual super-Poincar\'e algebra $\fp_Q$ of the minimal twist. Such a choice realizes the maximal twist as a further twist of the minimal twist. In the following, we sketch how the residual super-Poincar\'e algebra of the maximal twist arises as $(\fp_Q)_{Q_{\max}}$.

The choice of $Q_{\max}$ fixes a decomposition of the holomorphic and antiholomorphic directions in the minimal twist as
$
 L = L_{\max} \oplus V_3 $ and $L^\vee = L^\vee_{\max} \oplus V_3^\vee$.
Here, $V_3$ and $V_3^\vee$ are the additional directions rendered topological by $Q_{\max}$; together with the one direction that was already topological in the minimal twist, these form the seven topological directions of the maximal twist.

We can decompose the minimal residual super-Poincar\'e algebra under $\mathfrak{sl}(2) \times \mathfrak{sl}(3)$, where $L_{\max}$ and $V_3$ are the fundamental representations of $\mathfrak{sl}(2)$ and $\mathfrak{sl}(3)$ respectively. For the dg Lie algebra~${(\fp_Q, [Q_{\max}, -])}$, we find the following:
\[
 \hspace*{14.5mm}\begin{tikzcd}[row sep = tiny]
 \tikzmark{1} L_{\max}^\vee \\
 \tikzmark{2} V_3^\vee \\
 \tikzmark{3} \wedge^2 L_{\max}^\vee \\
 L_{\max}^\vee \otimes V_3^\vee & & L_{\max} \\
 \tikzmark{4} \wedge^2 V_3^\vee & \wedge^2 L_{\max} \otimes V_3 \arrow[r] & V_3\\
 \tikzmark{5} \mathfrak{gl}(1) \arrow[r] & \wedge^3 V_3
 \\
 L_{\max} \otimes V_3^\vee \arrow[r] & L_{\max} \otimes V_3^\vee\\
 L_{\max}^\vee \otimes V_3\\
 \mathfrak{sl}(3) \\
 \tikzmark{6} \mathfrak{sl}(2)
 \end{tikzcd}
 \begin{tikzpicture}[overlay,remember picture]
 \draw[decorate,decoration={brace}]
 ( $ (pic cs:2) -(0.25, 3pt) $ ) --
 ( $ (pic cs:1) +(-0.1, 10pt) $ )
 node[midway, left=5pt] {$L^\vee$};

 \draw[decorate,decoration={brace}]
 ( $ (pic cs:4) -(0.6, 3pt) $ ) --
 ( $ (pic cs:3) +(-0.45, 10pt) $ )
 node[midway, left=10pt] {$\wedge^2 L^\vee$};

 \draw[decorate,decoration={brace}]
 ( $ (pic cs:6) -(0.8, 3pt) $ ) --
 ( $ (pic cs:5) +(-0.8, 10pt) $ )
 node[midway, left=10pt] {$\mathfrak{sl}(5)$};
 \end{tikzpicture}
\]
Note that the cohomology in degree zero can be organized into the four pieces appearing in degree zero of the residual super-Poincar\'e algebra of the maximal twist. First, using the identification~$\wedge^2 V_3^\vee \cong V_3$, we have
$
 \mathfrak{sl}(3) \oplus V_3 \oplus V_3^\vee
$
which is the decomposition of the adjoint representation of $\fg_2$ under it $\mathfrak{sl}(3)$-subalgebra. Second, we identify
$
 L^\vee_{\max} \otimes \big(V_3 \oplus V_3^\vee \oplus \textbf{1}^0 \big)
$
with $V_7 \otimes \textbf{2}^{-1}$ and $\wedge^2 L_{\max}^\vee$ with $\textbf{1}^{-2}$. Finally, the identification of $\mathfrak{sl}(2)$ and the holomorphic translations as symmetries of the maximal twist is obvious.

\subsubsection[Homotopy transfer of the L\_infty structure]{Homotopy transfer of the $\boldsymbol{ L_\infty}$ structure}
The four-ary bracket on $(\fp_Q)_{Q_{\max}}$ arises as a combination of the brackets on $\fp_Q$ via homotopy transfer. One contribution comes from composing the four-ary bracket of $\fp_Q$ with the inclusion and projection in the obvious way. In addition, the three-ary and binary brackets contribute with the following two diagrams
\begin{equation*}
\begin{tikzpicture}
\begin{feynman}
\vertex at (-2,-0.5) {$\mu_4 \ = $};
\vertex(a) at (-1,1) {$i$};
\vertex(b) at (-1,0) {$i$};
\vertex(c) at (-1,-1) {$i$};
\vertex(c2) at (-1,-2) {$i$};
\vertex(d) at (0,0);
\vertex(e) at (1,0);
\vertex(e2) at (2,0){$p$};
\diagram* {(a)--(d), (b)--(d), (d)--[edge label = $h$](e), (c)--(d), (c2)--(e), (e)--(e2)};
\vertex(a') at (5,1) {$i$};
\vertex(b') at (5,0) {$i$};
\vertex(c') at (5,-1) {$i$};
\vertex(c2') at (5,-2) {$i$};
\vertex(d') at (6,0.5);
\vertex(e') at (7,0);
\vertex(e2') at (8,0){$p$.};
\diagram*{(a')--(d'), (b')--(d'), (d')--[edge label = $h$](e'), (c')--(e'), (c2')--(e'), (e')--(e2')};
\vertex at (3.4,-0.5) {$+$};
\end{feynman}
\end{tikzpicture}
\end{equation*}
Together, these terms give rise to the four-ary bracket in the maximal twist described in~Section~\ref{sec: max-ht}.

\subsubsection{Summary}
Similarly to the maximal twist, we have seen that higher brackets turn the residual super-Poincar\'e algebra into an $L_\infty$ algebra that is also represented on the minimally twisted theory in a non-strict way. Several directions here could be of interest for future work. First, our analysis in this paper is purely classical. It would be interesting to see how these higher operations behave under quantization. Since these actions are inner, the structure of quantum corrections are governed by BV-BRST anomalies of the theory. We anticipate that there are no anomalies in the maximal twist on flat space due to the results of \cite{williamswang} for holomorphic-topological theories with at least two topological directions (see also the conjectured formality of holomorphic-topological colored operads of \cite{Gaiotto:2024gii}). The minimal twist is less immediate: the obstruction to quantization to all orders in perturbation theory is controlled by the twelfth Gelfand--Fuchs cohomology of~$E(5|10)$ -- a group which remains unknown. However, we note that a similar formality theorem at the perturbative level for 3d twisted theories with a single topological direction was recently established in~\cite{Dimofte:2025oqf}. Finally, $L_\infty$ extensions of super-Poincar\'e algebras of a different flavor were studied in the context of the brane scan, which are meant to codify how spacetime super-Poincar\'e symmetries are corrected in the presence of extended objects in the theory~\cite{FSS}. These extensions also appear to play a crucial role in twisted supergravity, where the inclusion of such extensions is necessary for the map realizing the inner action of twisted supertranslations on the twisted fields to be a Lie map: see~\cite{CostelloLi, RSW} for examples in type IIB and eleven-dimensional supergravity respectively. We work out the relations between the higher symmetries constructed in this paper and brane extensions for twisted theories in our forthcoming work~\cite{twistedBranes}.

\appendix

\section{Additional details on some equivariant decompositions} \label{ap: details}

\subsection[Decomposition of the Lie bracket under G\_2 times SU(2)]{Decomposition of the Lie bracket under $\boldsymbol{ G_2 \times \SU(2)}$}
In the following, we decompose the Lie bracket in the eleven-dimensional $\cN=1$ super-Poincar\'e algebra $\fp$ under $G_2 \times \SU(2)$. These results allow us to evaluate all diagrams in the homotopy transfer calculation giving rise to the four-ary bracket described in~Section~\ref{p: maxtransfer}.

\subsubsection[so(V) times so(V) longrightarrow so(V)]{$\boldsymbol{\lie{so}(V) \times \lie{so}(V) \longrightarrow \lie{so}(V)}$}
Recall that $\lie{so}(V) \cong \wedge^2 V$ decomposes into contributions of three different types, corresponding to terms in $\lie{so}(V_7) \cong \wedge^2 V_7$, $\lie{so}(V_4) \cong \wedge^2 V_4$, and mixed terms in $V_7 \otimes V_4$.

For convenience, we also recall the $G_2 \times \SU(2) \times \U(1)$-equivariant decomposition of these components
\begin{gather*}
 \wedge^2 V_7 \cong \fg_2 \oplus V_7 ,\qquad\!
 V_7 \otimes V_4 \cong (V_7 \otimes \textbf{2}^1) \oplus \big(V_7 \otimes \textbf{2}^{-1}\big) ,\qquad\!
 \wedge^2 V_4 \cong \lie{sl}(2) \oplus \textbf{1}^1 \oplus \textbf{1}^0 \oplus \textbf{1}^{-1}.
\end{gather*}
With respect to this decomposition, the bracket has the following restrictions:
\begin{itemize}\itemsep=0pt
 \item[]\underline{$\lie{so}(V_7) \times \so(V) \longrightarrow \lie{so}(V)$}
 \begin{itemize}\itemsep=0pt
 \item[--] $\fg_2$ is a subalgebra, $[\fg_2 , \fg_2] \subset \fg_2$ and acts on all $\fg_2$-representations, i.e., $[\fg_2, V_7] \subset V_7$, $\big[\fg_2, V_7 \otimes \textbf{2}^{\pm 1}\big] \subset V_7 \otimes \textbf{2}^{\pm 1}$.
 \item[--] $V_7$ contains those rotations of $\R^7$ that do not respect the $G_2$-structure; we have $[V_7,V_7] \subset \fg_2 \oplus V_7$ and $\big[V_7, V_7 \otimes \textbf{2}^{\pm 1}\big] \subset V_7 \otimes \textbf{2}^{\pm 1}$.
 \end{itemize}
 \item[]\underline{$(V_7 \otimes V_4) \times \so(V) \longrightarrow \lie{so}(V)$}
 \begin{itemize}\itemsep=0pt
 \item[--] $\big[V_7 \otimes \textbf{2}^{1}, V_7 \otimes \textbf{2}^{-1}\big] \subset \fg_2 \oplus V_7 \oplus \lie{sl}(2) \oplus \textbf{1}^0$.
 \item[--] $\big[V_7 \otimes \textbf{2}^{\pm 1}, V_7 \otimes \textbf{2}^{\pm 1}\big] \subset \textbf{1}^{\pm 2}$.
 \end{itemize}
 \item[] \underline{$\lie{so}(V_4) \times \so(V) \longrightarrow \lie{so}(V)$}
 \begin{itemize}\itemsep=0pt
 \item[--] $[\lie{sl}(2), \lie{sl}(2)] \subset \lie{sl}(2)$ and $\lie{sl}(2)$ acts on all $\lie{sl}(2)$-representations.
 \item[--] $\big[\textbf{1}^2,\textbf{1}^{-2}\big] \subset \textbf{1}^0$, $\big[\textbf{1}^{\pm 2},\textbf{1}^{0}\big] \subset \textbf{1}^{\pm 2}$.
 \item[--] $\big[\textbf{1}^{\pm 2} , V_7 \otimes \textbf{2}^{\pm 1}\big] \subset V_7 \otimes \textbf{2}^{\mp 1}$, $\big[\textbf{1}^{0} , V_7 \otimes \textbf{2}^{\pm 1}\big] \subset V_7 \otimes \textbf{2}^{\pm 1}$.
 \end{itemize}
\end{itemize}

\subsubsection[so(V) times S longrightarrow S]{$\boldsymbol{\lie{so}(V) \times S \longrightarrow S}$}
Recall the decomposition
$
 S \cong (\textbf{1}_{G_2} \oplus V_7) \otimes \big(\textbf{2}^0 \oplus \textbf{1}^1 \oplus \textbf{1}^{-1}\big)$.
Here, the trivial representation $\textbf{1}_{G_2}$ corresponds to the spinor compatible with the $G_2$-structure on $\R^7$:
\begin{itemize}\itemsep=0pt
 \item[] \underline{$\lie{so}(V_7) \times S \longrightarrow S$}
 \begin{itemize}\itemsep=0pt
\item[--] $\fg_2$ acts according to representations, i.e., $[\fg_2 , \textbf{1}_{G_2} \otimes R] = 0$ and $[\fg_2 , \textbf{1}_{G_2} \otimes R] = R$ for all $\SU(2) \times \U(1)$-representations $R$.
\item[--] $V_7$ consists of those rotations that do not respect the spinor specified by the $G_2$-structure: $[V_7, \textbf{1}_{G_2} \otimes R] \subset V_7 \otimes R$ and $[V_7, V_7 \otimes R] \subset (\textbf{1}_{G_2} \oplus V_7) \otimes R$.
 \end{itemize}
 \item[] \underline{$(V_7 \otimes V_4) \times S \longrightarrow S$}
 \begin{itemize}\itemsep=0pt
\item[--] A mixed element $v \otimes w \in V_7 \otimes V_4$ acts on a spinor $\psi \otimes \sigma \in S_7 \otimes S_4$ through Clifford multiplication $(v \otimes w) \cdot (\psi \otimes \sigma) = (v \cdot \psi) \otimes (w \cdot \sigma)$. Clifford multiplication on the first factor maps $V_7 \otimes \textbf{1}_{G_2} \longrightarrow V_7$ and $V_7 \otimes V_7 \longrightarrow \textbf{1}_{G_2} \oplus V_7$; on the second factor, we have~${\textbf{2}^{\pm 1} \otimes \textbf{2}^0 \longrightarrow \textbf{1}^{\pm 1}}$ and $\textbf{2}^{\pm 1} \otimes \textbf{1}^{\pm 1} \longrightarrow \textbf{2}^{0}$.
 \end{itemize}
 \item[]\underline{$\lie{so}(V_4) \times S \longrightarrow S$}
 \begin{itemize}\itemsep=0pt
 \item[--] \lie{sl}(2) acts according to representation.
 \item[--] $\textbf{1}^q$ for $q = 2,0,-2$ acts by shifting the $\U(1)$ weights by $q$ where possible..=
 \end{itemize}
\end{itemize}

\subsubsection[so(V) times V longrightarrow V]{$\boldsymbol{\lie{so}(V) \times V \longrightarrow V}$}
\begin{itemize}\itemsep=0pt
 \item[]\underline{$\lie{so}(V_7) \times V \longrightarrow V$}
 \begin{itemize}\itemsep=0pt
 \item[--] $[\fg_2, V_7] \subset V_7$ and $[V_7 , V_7] \subset V_7$.
 \end{itemize}
 \item[]\underline{$(V_7 \otimes V_4) \times V \longrightarrow V$}
 \begin{itemize}\itemsep=0pt
\item[--] $\big[V_7 \otimes \textbf{2}^{\pm 1} , V_7\big] \subset \textbf{2}^{\pm 1}$ and $\big[V_7 \otimes \textbf{2}^{\pm 1} , \textbf{2}^{\mp 1}\big] \subset V_7$.
 \end{itemize}
 \item[]\underline{$\lie{so}(V_4) \times V \longrightarrow V$}
\begin{itemize}\itemsep=0pt
\item[--] \lie{sl}(2) acts according to representation.
\item[--] $\big[\textbf{1}^0 , \textbf{2}^{\pm 1}\big] \subset \textbf{2}^{\pm 1}$ and $\big[\textbf{1}^{\pm 2} , \textbf{2}^{\mp 1}\big] \subset \textbf{2}^{\pm 1}$.
 \end{itemize}
\end{itemize}

\subsection{Decomposition of vector fields}
Recall that \lie{so}(11) acts on $\R^{11}$ via vector fields
\begin{equation}\label{eq: so-action}
 x_\mu \partial_\nu - x_\nu \partial_\mu .
\end{equation}
In the following, we spell out how these vector fields relate to the decomposition of $\so(11)$ both in the maximal and minimal twists.

\subsubsection{Maximal twist}
Identifying $\R^{11} \cong \R^7 \times \C^2$, we define coordinates by $x^a = x^\mu$ for $\mu = 1,\dots, 7$ and
\begin{gather*}
 z^1 = x^{8} + i x^{9} , \qquad z^2 = x^{10} + i x^{11},\qquad
 \bar{z}_1 = x_{8} - i x_{9} , \qquad \bar{z}_2 = x_{10} - i x_{11}.
\end{gather*}
The vector fields~\eqref{eq: so-action} then decompose as follows:
\begin{table}[!ht]\renewcommand{\arraystretch}{1.2}
\centering
\begin{tabular}{|l|l|l|}
\hline
Representation & Vector field & Remarks \\ \hline
\(\fg_2\) & \(A^{ab} (x_a \partial_b - x_b \partial_a)\) & \(A^{ab} \varphi_{abc} = 0\) \\
\hline
\(V_7\) & \(a^{ab} (x_a \partial_b - x_b \partial_a)\) & \(a_{ab} = v^c \varphi_{abc}, v \in V_7\) \\
 \hline
\(V_7 \otimes \textbf{2}^{-1}\) & \(z^i \partial_{x_a} - x_a \partial_{\bar{z}_i}\) & \\
\hline
\(V_7 \otimes \textbf{2}^{1}\) & \(\bar{z}_i \partial_{x_a} - x_a \partial_{z^i} \) & \\
\hline
 & \(z_1 \partial_{z_2} - \bar{z}_2 \partial_{\bar{z}_1}\) & \\
\(\lie{sl}(2)\) & \(z^1 \partial_{z^1} - z^2 \partial_{z^2} - (\bar{z}_1 \partial_{\bar{z}_1} - \bar{z}_2 \partial_{\bar{z}_2})\) & \\
 & \(z^2 \partial_{z^1} - \bar{z}_1 \partial_{\bar{z}_2}\) & \\
\hline
\(\textbf{1}^{0}\) & \(z^1 \partial_{z^1} + z^2 \partial_{z^2} - (\bar{z}_1 \partial_{\bar{z}_1} + \bar{z}_2 \partial_{\bar{z}_2})\) & \\
 \hline
\(\textbf{1}^{2}\) & \(\bar{z}_1 \partial_{z^2} - \bar{z}_2 \partial_{z^1}\) & \\
 \hline
\(\textbf{1}^{-2}\) & \(z^1 \partial_{\bar{z}_2} - z^2 \partial_{\bar{z}_1}\)& \\
\hline
\end{tabular}
\end{table}

\pagebreak

\subsubsection{Minimal twist}

Identifying $\R^{11} \cong \CC^5 \times \R$, we can introduce coordinates $\big(z^i, \bar{z}_i , t\big)$ with $i=1,\dots, 5$. Then the action of $\so(11)$ via vector fields decomposes as follows:
\begin{table}[!ht]\renewcommand{\arraystretch}{1.2}
\centering
\begin{tabular}{|l|l|}
\hline
Representation & Vector field \\ \hline
\( L^\vee \) & \( z^i \partial_t - t \partial_{\bar{z}_i} \) \\
\hline
\( \wedge^2 L^\vee \) & \( z^i \partial_{\bar{z}_j} - z^j \partial_{\bar{z}_i} \) \\ \hline
\( \mathfrak{sl}(5) \) & \( A_{i}^{\ j} \big(z^i \partial_{z^j} + \bar{z}_j \partial_{\bar{z}_i}\big) \qquad \text{with} \qquad A_{i}^{\ j} \in \mathfrak{sl}(5) \) \\
\hline
\( \wedge^2 L \) & \( \bar{z}_i \partial_{z^j} - \bar{z}_j \partial_{z^i} \) \\
\hline
\( L \) & \( \bar{z}_i \partial_t - t \partial_{z^i}\) \\
\hline
\( \mathfrak{gl}(1) \) & \(z^i \partial_{z^i} + \bar{z}_i \partial_{\bar{z}_i}\) \\
\hline
\end{tabular}
\end{table}

\subsection*{Acknowledgments}
We would like to thank K.~Costello, N.~Garner, and I.~Saberi for helpful discussions. In addition, we would like to thank the anonymous referees for useful suggestions. NP and FH are both supported by the DOE Early Career Research Program under award DE-SC0022924. NP is also supported by funds from the Department of Physics and the College of Arts \& Sciences at the University of Washington, and the Simons Foundation as part of the Simons Collaboration on Celestial Holography.

\pdfbookmark[1]{References}{ref}
\LastPageEnding


\begin{thebibliography}{99}
\footnotesize\itemsep=0pt

\bibitem{SecondaryOps}
Beem C., Ben-Zvi D., Bullimore M., Dimofte T., Neitzke A., Secondary products
 in supersymmetric field theory,
 \href{https://doi.org/10.1007/s00023-020-00888-3}{\textit{Ann. Henri
 Poincar\'e}} \textbf{21} (2020), 1235--1310,
 \href{http://arxiv.org/abs/1809.00009}{arXiv:1809.00009}.

\bibitem{Beem:2013sza}
Beem C., Lemos M., Liendo P., Peelaers W., Rastelli L., van Rees B.C., Infinite
 chiral symmetry in four dimensions,
 \href{https://doi.org/10.1007/s00220-014-2272-x}{\textit{Comm. Math. Phys.}}
 \textbf{336} (2015), 1359--1433,
 \href{http://arxiv.org/abs/1312.5344}{arXiv:1312.5344}.

\bibitem{BerkovitsSupermembrane}
Berkovits N., Towards covariant quantization of the supermembrane,
 \href{https://doi.org/10.1088/1126-6708/2002/09/051}{\textit{J.~High Energy
 Phys.}} \textbf{2002} (2002), no.~9, 051, 44~pages,
 \href{http://arxiv.org/abs/hep-th/0201151}{arXiv:hep-th/0201151}.

\bibitem{Bomans:2023mkd}
Bomans P., Wu J., Unravelling the holomorphic twist: central charges,
 \href{https://doi.org/10.1007/s00220-024-05167-4}{\textit{Comm. Math. Phys.}}
 \textbf{405} (2024), 290, 49~pages,
 \href{http://arxiv.org/abs/2311.04304}{arXiv:2311.04304}.

\bibitem{Budzik:2023xbr}
Budzik K., Gaiotto D., Kulp J., Williams B.R., Wu J., Yu M., Semi-chiral
 operators in 4d~{$\mathcal{N} = 1$} gauge theories,
 \href{https://doi.org/10.1007/jhep05(2024)245}{\textit{J.~High Energy Phys.}}
 \textbf{2024} (2024), no.~5, 245, 69~pages,
 \href{http://arxiv.org/abs/2306.01039}{arXiv:2306.01039}.

\bibitem{Budzik:2022mpd}
Budzik K., Gaiotto D., Kulp J., Wu J., Yu M., Feynman diagrams in
 four-dimensional holomorphic theories and the {O}peratope,
 \href{https://doi.org/10.1007/jhep07(2023)127}{\textit{J.~High Energy Phys.}}
 \textbf{2023} (2023), no.~7, 127, 38~pages,
 \href{http://arxiv.org/abs/2207.14321}{arXiv:2207.14321}.

\bibitem{Ced-11d}
Cederwall M., {$D=11$} supergravity with manifest supersymmetry,
 \href{https://doi.org/10.1142/S0217732310034407}{\textit{Modern Phys.
 Lett.~A}} \textbf{25} (2010), 3201--3212,
 \href{http://arxiv.org/abs/1001.0112}{arXiv:1001.0112}.

\bibitem{Ced-SL5}
Cederwall M., {${\rm SL}(5)$} supersymmetry,
 \href{https://doi.org/10.1002/prop.202100116}{\textit{Fortschr. Phys.}}
 \textbf{69} (2021), 2100116, 5~pages,
 \href{http://arxiv.org/abs/2107.09037}{arXiv:2107.09037}.

\bibitem{Cederwall_2019}
Cederwall M., Palmkvist J., {$L_\infty$} algebras for extended geometry,
 \href{https://doi.org/10.1088/1742-6596/1194/1/012021}{\textit{J.~Phys. Conf.
 Ser.}} \textbf{1194} (2019), 012021, 9~pages.

\bibitem{CostelloMtheory1}
Costello K., Holography and Koszul duality: the example of the~$M2$ brane,
 \href{http://arxiv.org/abs/1705.02500}{arXiv:1705.02500}.

\bibitem{CostelloMtheory2}
Costello K., M-theory in the omega-background and 5-dimensional non-commutative
 gauge theory, \href{http://arxiv.org/abs/1610.04144}{arXiv:1610.04144}.

\bibitem{CostelloHol}
Costello K., Notes on supersymmetric and holomorphic field theories in
 dimensions~2 and~4,
 \href{https://doi.org/10.4310/PAMQ.2013.v9.n1.a3}{\textit{Pure Appl.
 Math.~Q.}} \textbf{9} (2013), 73--165,
 \href{http://arxiv.org/abs/1111.4234}{arXiv:1111.4234}.

\bibitem{Costello:2018zrm}
Costello K., Gaiotto D., Twisted holography,
 \href{https://doi.org/10.1007/JHEP01(2025)087}{\textit{J.~High Energy Phys.}}
 \textbf{2025} (2025), no.~01, 087, 88~pages,
 \href{http://arxiv.org/abs/1812.09257}{arXiv:1812.09257}.

\bibitem{CG1}
Costello K., Gwilliam O., Factorization algebras in quantum field theory.
 {V}ol.~1, \textit{New Math. Monogr.}, Vol.~31,
 \href{https://doi.org/10.1017/9781316678626}{Cambridge University Press},
 Cambridge, 2017.

\bibitem{CG2}
Costello K., Gwilliam O., Factorization algebras in quantum field theory.
 {V}ol.~2, \textit{New Math. Monogr.}, Vol.~41,
 \href{https://doi.org/10.1017/9781316678664}{Cambridge University Press},
 Cambridge, 2021.

\bibitem{CostelloLi}
Costello K., Li S., Twisted supergravity and its quantization,
 \href{http://arxiv.org/abs/1606.00365}{arXiv:1606.00365}.

\bibitem{Costello:2020jbh}
Costello K., Paquette N.M., Twisted supergravity and {K}oszul duality: a case
 study in~{$\rm AdS_3$},
 \href{https://doi.org/10.1007/s00220-021-04065-3}{\textit{Comm. Math. Phys.}}
 \textbf{384} (2021), 279--339,
 \href{http://arxiv.org/abs/2001.02177}{arXiv:2001.02177}.

\bibitem{CJS}
Cremmer E., Julia B., Scherk J., Supergravity theory in eleven-dimensions,
 \href{https://doi.org/10.1016/0370-2693(78)90894-8}{\textit{Phys. Lett.~B}}
 \textbf{76} (1978), 409--412.

\bibitem{Cushing:2023rha}
Cushing J., Moore G.W., Ro{\v{c}}ek M., Saxena V., Superconformal gravity and
 the topology of diffeomorphism groups,
 \href{http://arxiv.org/abs/2311.08394}{arXiv:2311.08394}.

\bibitem{EvidenceG2}
Del~Zotto M., Oh J., Zhou Y., Evidence for an algebra of~{$G_2$} instantons,
 \href{https://doi.org/10.1007/jhep08(2022)214}{\textit{J.~High Energy Phys.}}
 \textbf{2022} (2022), no.~8, 214, 57~pages,
 \href{http://arxiv.org/abs/2109.01110}{arXiv:2109.01110}.

\bibitem{Dimofte:2025oqf}
Dimofte T., Niu W., Py V., Line operators in 3d holomorphic qft: meromorphic
 tensor categories and dg-shifted {Y}angians,
 \href{http://arxiv.org/abs/2508.11749}{arXiv:2508.11749}.

\bibitem{MaxTwist}
Eager R., Hahner F., Maximally twisted eleven-dimensional supergravity,
 \href{https://doi.org/10.1007/s00220-022-04516-5}{\textit{Comm. Math. Phys.}}
 \textbf{398} (2023), 59--88,
 \href{http://arxiv.org/abs/2106.15640}{arXiv:2106.15640}.

\bibitem{NV}
Eager R., Saberi I., Walcher J., Nilpotence varieties,
 \href{https://doi.org/10.1007/s00023-020-01007-y}{\textit{Ann. Henri
 Poincar\'e}} \textbf{22} (2021), 1319--1376,
 \href{http://arxiv.org/abs/1807.03766}{arXiv:1807.03766}.

\bibitem{sucotwist}
Elliott C., Gwilliam O., Lotito M., Twists of superconformal algebra,
 \href{http://arxiv.org/abs/2403.19753}{arXiv:2403.19753}.

\bibitem{ElliottSafronov}
Elliott C., Safronov P., Topological twists of supersymmetric algebras of
 observables, \href{https://doi.org/10.1007/s00220-019-03393-9}{\textit{Comm.
 Math. Phys.}} \textbf{371} (2019), 727--786,
 \href{http://arxiv.org/abs/1805.10806}{arXiv:1805.10806}.

\bibitem{Festuccia:2011ws}
Festuccia G., Seiberg N., Rigid supersymmetric theories in curved superspace,
 \href{https://doi.org/10.1007/JHEP06(2011)114}{\textit{J.~High Energy Phys.}}
 \textbf{2011} (2011), no.~6, 114, 23~pages,
 \href{http://arxiv.org/abs/1105.0689}{arXiv:1105.0689}.

\bibitem{FSS}
Fiorenza D., Sati H., Schreiber U., Super-{L}ie {$n$}-algebra extensions,
 higher {WZW} models and super-{$p$}-branes with tensor multiplet fields,
 \href{https://doi.org/10.1142/S0219887815500188}{\textit{Int.~J. Geom.
 Methods Mod. Phys.}} \textbf{12} (2015), 1550018, 35~pages.

\bibitem{Gaberdiel:1997ia}
Gaberdiel M.R., Zwiebach B., Tensor constructions of open string theories.~{I}.
 {F}oundations,
 \href{https://doi.org/10.1016/S0550-3213(97)00580-4}{\textit{Nuclear
 Phys.~B}} \textbf{505} (1997), 569--624,
 \href{http://arxiv.org/abs/hep-th/9705038}{arXiv:hep-th/9705038}.

\bibitem{Gaiotto:2024gii}
Gaiotto D., Kulp J., Wu J., Higher operations in perturbation theory,
 \href{https://doi.org/10.1007/jhep05(2025)230}{\textit{J.~High Energy Phys.}}
 \textbf{2025} (2025), no.~5, 230, 75~pages,
 \href{http://arxiv.org/abs/2403.13049}{arXiv:2403.13049}.

\bibitem{Garner:2022its}
Garner N., Paquette N.M., Mathematics of String Dualities,
 \href{https://doi.org/10.22323/1.403.0007}{\textit{Proc. of Sci.}}
 \textbf{403} (2023), PoS(TASI2021)007, 85~pages,
 \href{http://arxiv.org/abs/2204.01914}{arXiv:2204.01914}.

\bibitem{Garner:2023wrc}
Garner N., Raghavendran S., Williams B.R., Enhanced symmetries in minimally
 twisted three-dimensional supersymmetric theories,
 \href{https://doi.org/10.4310/atmp.250929221027}{\textit{Adv. Theor. Math.
 Phys.}} \textbf{29} (2025), 815--862,
 \href{http://arxiv.org/abs/2310.08516}{arXiv:2310.08516}.

\bibitem{garnerhiggscoulomb}
Garner N., Raghavendran S., Williams B.R., Higgs and {C}oulomb branches from
 superconformal raviolo vertex algebras,
 \href{https://doi.org/10.1016/j.aim.2025.110566}{\textit{Adv. Math.}}
 \textbf{482} (2025), 110566, 44~pages,
 \href{http://arxiv.org/abs/2310.08524}{arXiv:2310.08524}.

\bibitem{malgebra}
Giotopoulos G., Sati H., Schreiber U., The {M}-algebra completes the hierarchy
 of super-exceptional tangent spaces,
 \href{https://doi.org/10.1016/j.physletb.2024.139199}{\textit{Phys. Lett.~B}}
 \textbf{860} (2025), 139199, 10~pages,
 \href{http://arxiv.org/abs/2024.13919}{arXiv:2024.13919}.

\bibitem{BPS-SS}
Gukov S., Nawata S., Saberi I., Sto\v{s}i\'c M., Su\l~kowski P., Sequencing
 {BPS} spectra, \href{https://doi.org/10.1007/JHEP03(2016)004}{\textit{J.~High
 Energy Phys.}} \textbf{2016} (2016), no.~3, 004, 160~pages,
 \href{http://arxiv.org/abs/1512.07883}{arXiv:1512.07883}.

\bibitem{twistedBranes}
Hahner F., Paquette N.M., Raghavendran S., Saberi I., {i}n
 preparation.

\bibitem{LSCA}
Hahner F., Raghavendran S., Saberi I., Williams B.R., Local superconformal
 algebras, \href{http://arxiv.org/abs/2410.08176}{arXiv:2410.08176}.

\bibitem{CY2}
Hahner F., Saberi I., Eleven-dimensional supergravity as a {C}alabi--{Y}au
 twofold, \href{https://doi.org/10.1007/s00029-025-01024-x}{\textit{Selecta
 Math.~(N.S.)}} \textbf{31} (2025), 38, 34~pages,
 \href{http://arxiv.org/abs/2304.12371}{arXiv:2304.12371}.

\bibitem{MaldacenaHerderschee}
Herderschee A., Maldacena J., Three point amplitudes in matrix theory,
 \href{https://doi.org/10.1088/1751-8121/ad389b}{\textit{J.~Phys.~A}}
 \textbf{57} (2024), 165401, 21~pages,
 \href{http://arxiv.org/abs/2312.12592}{arXiv:2312.12592}.

\bibitem{Hull_2007}
Hull C.M., Generalised geometry for {M}-theory,
 \href{https://doi.org/10.1088/1126-6708/2007/07/079}{\textit{J.~High Energy
 Phys.}} \textbf{2007} (2007), no.~7, 079, 31~pages,
 \href{http://arxiv.org/abs/hep-th/0701203}{arXiv:hep-th/0701203}.

\bibitem{JohansenHet}
Johansen A., Twisting of~{$N=1$} {SUSY} gauge theories and heterotic
 topological theories,
 \href{https://doi.org/10.1142/S0217751X9500200X}{\textit{Internat.~J. Modern
 Phys.~A}} \textbf{10} (1995), 4325--4357.

\bibitem{JKY}
Jonsson D.S.H., Kim H., Young C.A.S., Homotopy representations of extended
 holomorphic symmetry in holomorphic twists,
 \href{https://doi.org/10.1016/j.geomphys.2025.105632}{\textit{J.~Geom.
 Phys.}} \textbf{217} (2025), 105632, 16~pages,
 \href{http://arxiv.org/abs/2408.00704}{arXiv:2408.00704}.

\bibitem{Jurco:2020yyu}
Jur\v{c}o B., Kim H., Macrelli T., Saemann C., Wolf M., Perturbative quantum
 field theory and homotopy algebras,
 \href{https://doi.org/10.22323/1.376.0199}{\textit{Proc. of Sci.}}
 \textbf{376} (2020), PoS(CORFU2019)199, 25~pages,
 \href{http://arxiv.org/abs/2002.11168}{arXiv:2002.11168}.

\bibitem{Jurco:2018sby}
Jur\v{c}o B., Raspollini L., S\"amann C., Wolf M., {$L_\infty$}-algebras of
 classical field theories and the {B}atalin--{V}ilkovisky formalism,
 \href{https://doi.org/10.1002/prop.201900025}{\textit{Fortschr. Phys.}}
 \textbf{67} (2019), 1900025, 60~pages,
 \href{http://arxiv.org/abs/1809.09899}{arXiv:1809.09899}.

\bibitem{LadaMarkl}
Lada T., Markl M., Strongly homotopy {L}ie algebras,
 \href{https://doi.org/10.1080/00927879508825335}{\textit{Comm. Algebra}}
 \textbf{23} (1995), 2147--2161,
 \href{http://arxiv.org/abs/hep-th/9406095}{arXiv:hep-th/9406095}.

\bibitem{LodayVallette}
Loday J.-L., Vallette B., Algebraic operads, \textit{Grundlehren Math. Wiss.},
 Vol.~346, \href{https://doi.org/10.1007/978-3-642-30362-3}{Springer},
 Heidelberg, 2012.

\bibitem{perturbiner}
Lopez-Arcos C., V\'elez A.Q., {$L_\infty$}-algebras and the perturbiner
 expansion, \href{https://doi.org/10.1007/jhep11(2019)010}{\textit{J.~High
 Energy Phys.}} \textbf{2019} (2019), no.~11, 010, 31~pages,
 \href{http://arxiv.org/abs/1907.12154}{arXiv:1907.12154}.

\bibitem{Scattering-recursion}
Macrelli T., S\"amann C., Wolf M., Scattering amplitude recursion relations in
 {B}atalin--{V}ilkovisky-quantizable theories,
 \href{https://doi.org/10.1103/physrevd.100.045017}{\textit{Phys. Rev.~D}}
 \textbf{100} (2019), 045017, 22~pages,
 \href{http://arxiv.org/abs/1903.05713}{arXiv:1903.05713}.

\bibitem{nikitathesis}
Nekrassov N., Four-dimensional holomorphic theories, Ph.D.~Thesis, {P}rinceton
 University, 1996, \url{https://www.proquest.com/docview/304261334}.

\bibitem{RSW}
Raghavendran S., Saberi I., Williams B.R., Twisted eleven-dimensional
 supergravity, \href{https://doi.org/10.1007/s00220-023-04745-2}{\textit{Comm.
 Math. Phys.}} \textbf{402} (2023), 1103--1166,
 \href{http://arxiv.org/abs/2111.03049}{arXiv:2111.03049}.

\bibitem{RW}
Raghavendran S., Williams B.R., A holographic approach to the six-dimensional
 superconformal index,
 \href{http://arxiv.org/abs/2210.07910}{arXiv:2210.07910}.

\bibitem{SWchar}
Saberi I., Williams B.R., Twisted characters and holomorphic symmetries,
 \href{https://doi.org/10.1007/s11005-020-01319-4}{\textit{Lett. Math. Phys.}}
 \textbf{110} (2020), 2779--2853,
 \href{http://arxiv.org/abs/1906.04221}{arXiv:1906.04221}.

\bibitem{SCA}
Saberi I., Williams B.R., Superconformal algebras and holomorphic field
 theories, \href{https://doi.org/10.1007/s00023-022-01224-7}{\textit{Ann.
 Henri Poincar\'e}} \textbf{24} (2023), 541--604,
 \href{http://arxiv.org/abs/1910.04120}{arXiv:1910.04120}.

\bibitem{spinortwist}
Saberi I., Williams B.R., Twisting pure spinor superfields, with applications
 to supergravity,
 \href{https://doi.org/10.4310/pamq.2024.v20.n2.a2}{\textit{Pure Appl.
 Math.~Q.}} \textbf{20} (2024), 645--701,
 \href{http://arxiv.org/abs/2106.15639}{arXiv:2106.15639}.

\bibitem{smithwaldram}
Smith G.R., Waldram D., M-theory moduli from exceptional complex structures,
 \href{https://doi.org/10.1007/jhep08(2023)022}{\textit{J.~High Energy Phys.}}
 \textbf{2023} (2023), no.~8, 22, 36~pages,
 \href{http://arxiv.org/abs/2211.09517}{arXiv:2211.09517}.

\bibitem{Stelzig}
Stelzig J., On the structure of double complexes,
 \href{https://doi.org/10.1112/jlms.12453}{\textit{J.~Lond. Math. Soc.~(2)}}
 \textbf{104} (2021), 956--988,
 \href{http://arxiv.org/abs/1812.00865}{arXiv:1812.00865}.

\bibitem{williamswang}
Wang M., Williams B.R., Factorization algebras from topological-holomorphic
 field theories, \href{http://arxiv.org/abs/2407.08667}{arXiv:2407.08667}.

\bibitem{Zeng:2023qqp}
Zeng K., Twisted holography and celestial holography from boundary chiral
 algebra, \href{https://doi.org/10.1007/s00220-023-04917-0}{\textit{Comm.
 Math. Phys.}} \textbf{405} (2024), 19, 109~pages,
 \href{http://arxiv.org/abs/2302.06693}{arXiv:2302.06693}.

\bibitem{Zwiebach:1992ie}
Zwiebach B., Closed string field theory: quantum action and the
 {B}atalin--{V}ilkovisky master equation,
 \href{https://doi.org/10.1016/0550-3213(93)90388-6}{\textit{Nuclear Phys.~B}}
 \textbf{390} (1993), 33--152,
 \href{http://arxiv.org/abs/hep-th/9206084}{arXiv:hep-th/9206084}.

\end{thebibliography}
\end{document}